\documentclass[fleqn,usenatbib]{mnras}

\usepackage{newtxtext,newtxmath}

\usepackage[T1]{fontenc}
\usepackage{ae,aecompl}

\usepackage[dvipdfmx]{graphicx} 
\usepackage{amsmath}  
\usepackage{indentfirst}  
\usepackage{pict2e}
\usepackage{color}
\usepackage{bm}
\usepackage{tablefootnote}
\usepackage{pict2e}

\usepackage{ulem}

\newcommand{\msun}{M_{\odot}}

\newcommand{\msunyr}{M_\odot~{\rm yr}^{-1}}

\newcommand{\hii}{H{\sc ii} }

\title[Massive Very Metal-Poor Stars]{Formation of massive stars under protostellar radiation feedback: Very metal-poor stars}

\author[Fukushima et al.]{ Hajime Fukushima$^{1, 2, 4}$\thanks{E-mail:fukushima@ccs.tsukuba.ac.jp}, Takashi Hosokawa$^{2}$, Gen Chiaki$^{3}$, Kazuyuki Omukai$^{4}$,
\newauthor Naoki Yoshida$^{5, 6}$ and Rolf Kuiper$^{7}$
\\
$^{1}$Center for Computational Sciences, University of Tsukuba, Ten-nodai, 1-1-1 Tsukuba, Ibaraki 305-8577, Japan\\
$^{2}$Department of Physics, Kyoto University, Sakyo, Kyoto 606-8502, Japan\\
$^{3}$Center for Relativistic Astrophysics, School of Physics, Georgia Institute of Technology, Atlanta, GA 30332, USA\\
$^{4}$Astronomical Institute, Graduate School of Science, Tohoku University, Aoba, Sendai 980-8578, Japan\\
$^{5}$Department of Physics, School of Science, The University of Tokyo, 7-3-1 Hongo, Bunkyo, Tokyo 113-0033, Japan\\
$^{6}$Kavli Institute for the Physics and Mathematics of the Universe (WPI), UT Institute for Advanced Study, The University of Tokyo, Kashiwa, Chiba 277-8583, Japan\\
$^{7}$Institute of Astronomy and Astrophysics, University of T\"ubingen, Auf der Morgenstelle 10, D-72076 T\"ubingen, Germany
}

\date{Accepted XXX. Received YYY; in original form ZZZ}

\pubyear{2018}

\begin{document}
\label{firstpage}
\pagerange{\pageref{firstpage}--\pageref{lastpage}}
\maketitle

\begin{abstract}
We study the formation of very metal-poor stars under protostellar radiative feedback effect.  
We use cosmological simulations to identify low-mass dark matter halos and
star-forming gas clouds within them. 
We then follow protostar formation and the subsequent long-term mass accretion phase of over one million years using two-dimensional radiation-hydrodynamics simulations.
We show that the critical physical process that sets the final mass is formation and expansion of a bipolar H{\sc ii} region. The process is similar to the formation of massive primordial stars, but radiation pressure exerted on dust grains also contributes to halting the accretion flow in the low-metallicity case. We find that the net feedback effect in the case with  metallicity $Z = 10^{-2}~Z_\odot$ is {\it stronger} than in the case with $Z \sim 1~Z_\odot$. With decreasing metallicity, the radiation pressure effect becomes weaker, but photoionization heating of the circumstellar gas is more efficient owing to the reduced dust attenuation. 
In the case with $Z = 10^{-2}~Z_\odot$,
the central star grows as massive as
 200 solar-masses, similarly to
 the case of primordial star formation.
We conclude that metal-poor stars with a few hundred solar masses 
can be formed by gas accretion despite the strong radiative feedback.
\end{abstract}

\begin{keywords}
stars: Population II - stars: massive - stars: formation - cosmology: theory - accretion, accretion discs
\end{keywords}


\section{Introduction}\label{introduction}

Massive stars cause a variety of feedback effects to their surroundings.
Radiation and mechanical energy input from them
affect the structure and evolution of the interstellar medium, 
and regulate the galactic-scale star formation. 
Heavy elements synthesized in stellar interior are ejected by stellar wind and 
supernova explosions, to drive the galactic and cosmic chemical evolution. 
It is expected that massive stars play an important role in 
galaxy formation in the early universe through the feedback processes.


Formation of massive primordial stars has been studied theoretically by a number of authors.
The gravitational collapse of a chemically pristine gas cloud induced by $\rm H_2$ and HD molecular cooling \citep[e.g.,][]{2002Sci...295...93A, Bromm02,Y03,2006ApJ...652....6Y} results in the formation of a protostar at the center when the gas density exceeds $\sim 10^ {20}~{\rm cm}^{-3}$ \citep[e.g.,][]{1998ApJ...508..141O,2008Sci...321..669Y}. 
Although the newly-born protostar is initially very small, it rapidly grows in mass  
by accreting the gas from the surrounding envelope at a rate approximated as $\dot M_* \sim c_{\rm s}^3 /G \propto T^{3/2}$, where $c_{\rm s}$ and $T$ are the sound speed and temperature in the star-forming cloud. The typical temperature in the primordial case, $T \simeq 300$~K, yields a large accretion rate of $\dot M \sim 10^{-3}~\msunyr$;
a 10 $\msun$ star can be formed in just about ten thousand years.


The final mass of the forming star is set when the accretion is terminated.
Radiation from the protostar heats up the surrounding gas, exerts radiation force onto it, and eventually halt the accretion flow. 
Photoionization is the primary feedback effect in primordial star formation. 
Although photoionization heating was considered to be ineffective under the assumption of
spherical symmetry \citep[e.g.,][]{1971A&A....13..190L, 2002MNRAS.332...59O}, recent studies suggest that it is strong enough to terminate 
gas accretion in a realistic case with disk accretion \citep[][]{2008ApJ...681..771M, 2011Sci...334.1250H},

When a protostar is surrounded by a circumstellar disk, \hii regions expand
in the polar directions of the disk, where the density is relatively low, 
and then expand into the outer envelope. 
Consequently, the disk begins to be directly exposed to the stellar ionizing radiation, and is eventually destroyed by photoevaporation. 
The final stellar mass is typically large, and often exceeds a few tens solar-masses in many cases \citep[e.g.,][]{2016ApJ...824..119H,2016MNRAS.462.1307S}. 
Statistical studies using a suite of cosmological simulations show the stellar mass distribution 
extends broadly from 10 to 1000 $\msun$ \citep{2014ApJ...781...60H, 2015MNRAS.448..568H, 2014ApJ...792...32S}.


Very massive stars with $\sim 100~\msun$ are also found forming in the present-day universe, from
gas that contains heavy elements and dust grains \citep[e.g.,][]{ZY07}. 
Observations suggest that the very massive stars form via rapid mass accretion at a rate of $10^{-4}$ $-$ $10^{-2}~\msunyr$ \citep[e.g.,][]{2005A&A...442..949F, 2016A&A...585A.149W}. Such a rate is much higher than that expected from the typical gas temperature $T \sim 10$~K, and, by coincidence, is comparable to the theoretically expected rate in primordial star formation. 

In a metal-enriched gas, radiation pressure on dust grains is thought to be of primarily importance.
Theoretical studies suggest that radiation pressure is so strong that the mass accretion is halted before the stellar mass exceeds $\simeq 30~\msun$ in the case of spherical symmetry \citep[e.g.,][]{1971A&A....13..190L, 1974A&A....37..149K, 1977A&A....54..183Y,1987ApJ...319..850W}.
Interestingly, more recent studies that consider multi-dimensional effects show overall weakening of the feedback effect.
Two and three-dimensional radiation-hydrodynamics (RHD) simulations show consistently that the stellar mass growth via disk accretion continues well beyond $30~\msun$  \citep[e.g.,][]{Yorke2002, Yorke1999, Krumholz2009, Kuiper2010, Kuiper2012, Klassen2016, Rosen2016, Harries2017,2020A&A...635A..42M}. 
However, the photoionization effect has not been incorporated into these studies except for \citet{2019MNRAS.486.5171S}. 
The effect has been studied only as a minor or a secondary feedback process \citep[e.g.,][]{Keto03,2010ApJ...725..134P, 2011ApJ...729...72P}. 
Coupled, the nonlinear effects of radiation pressure and photoionization are investigated by \citet{2018A&A...616A.101K} 
using 2D RHD simulations. They find that the photoionization works against the radiation-pressure feedback, 
and the net effect is to promote more efficient stellar mass growth 
than only with the radiation-pressure effect. 

There are a number of differences in detailed mechanisms and in the strength of radiative feedback on massive star formation between the primordial ($Z = 0$) and the present-day ($Z \sim Z_\odot$) environments.
An important question then arises: is massive star formation 
in {\it metal-poor environments} similar to primordial/present-day star formation ? 
Low-metallicity star formation has been considered in the literature, but mostly 
with interest in formation of low-mass $(\lesssim 1~\msun)$ stars as a result of fragmentation \citep[e.g.,][]{2001MNRAS.328..969B,2003Natur.422..869S,2005ApJ...626..627O, 2016MNRAS.463.2781C}. 
{\it High-mass} stars may also form simultaneously 
in a low-metallicity gas cloud, and may play a vital role in the first galaxies. 
In order to study formation of massive metal-poor stars, it is important to follow the long-term evolution of the protostellar accretion. Radiative feedback from growing protostars is likely dominant to determine their final masses. \citet{2009ApJ...703.1810H} and \citet{2018MNRAS.473.4754F} evaluate the strength of feedback effects assuming spherical symmetry for $0 \leq Z \leq 1~Z_\odot$. They show that photoionization is effective in regulating the accretion flow in the range $Z < Z_{\rm cr} \simeq 10^{-3} -  10^{-2}~Z_\odot$ but radiation pressure becomes more important at $Z > Z_{\rm cr}$. \citet{2018ApJ...861...68T} use a semi-analytic method to conclude that the photoionization effect becomes less important with increasing $Z$ owing to decreasing dust attenuation of ionizing photons. 
Because of the complex interplay between photoionization and radiation pressure
as mentioned in the above, it should be ideal or even necessary to perform multi-dimensional radiation hydrodynamics simulations to draw a definite conclusion. 

In the present paper, we perform a suite of 2D RHD simulations to study the formation of massive, very metal-poor stars under radiative feedback. 
We first follow early run-away collapse of a metal-enriched cloud in a cosmological minihalo \citep[see also][]{2016MNRAS.463.2781C}, and then follow the subsequent long-term ($\sim$ Myr) accretion phase under radiative feedback. 
Our simulations include both the  radiation-pressure and photoionization effects as in \citet{2018A&A...616A.101K}, as well as relevant thermal and chemical processes in the very metal-poor gas. We show formation 
of very massive stars, and identify the dominant feedback mechanism that sets the final mass.

We organize the rest of the paper as follows. In Section \ref{NumericalMethod}, we first describe the numerical method and initial conditions of our simulations.
The main results are presented in Section \ref{Results}.
Finally, we discuss the implication of our results and summarize our calculations in Section \ref{matome}. We describe some details of our numerical method in Appendix \ref{chemapp}.

\section{Numerical Method} \label{NumericalMethod}

We first give an overview of our calculations as follows.
We follow in 3D the collapse of metal-enriched clouds found in cosmological simulations performed by \citet{2014ApJ...781...60H}, using a modified version of GADGET-2 \citep{2005MNRAS.364.1105S}. We use a detailed chemical network including reactions with gas-phase heavy elements and dust grains \citep{2016MNRAS.463.2781C}.
Next, we remap the 3D data of the final snapshot of the cloud collapse simulation onto 2D data,
to perform 2D axisymmetric RHD simulations that follow the evolution in the protostellar accretion phase.
We use a modified version of PLUTO 4.1 \citep{2007ApJS..170..228M} and the simplified chemical network. In the following subsections, we describe our numerical methods used for the early collapse phase (Sec.~\ref{initial_conditions}) and later accretion phase (Sec.~\ref{initial_conditions2}), separately.

\subsection{Numerical method: cloud collapse}
\label{initial_conditions}

We generate the initial conditions for our simulations using an output of the cosmological simulation performed by \citet{2014ApJ...781...60H}, who investigated statistical properties of a large number ($>100$) of primordial star-forming clouds in a simulation box of $1h^{-1}~{\rm Mpc}$ on a side.
The cosmological simulations were run with $N$-body/SPH solver GADGET-2  coupled with non-equilibrium chemistry network to follow the cloud formation in minihalos \citep[e.g.,][]{2006ApJ...652....6Y}. We select a cloud that collapses at redshift $\simeq 20.46$ over the short timescale $t_{\rm col} \simeq 2 ~{t_{\rm col, 0}}$, where $t_{\rm col, 0} = 1 / \sqrt{24 \pi G \rho}$. The cloud has one of the shortest collapse timescale among the 100 samples in simulations of \citet{2014ApJ...781...60H}. 
While \citet{2014ApJ...781...60H} has followed the cloud collapse to the birth of a protostar, we here use the earlier snapshot at the epoch when the central density is still $n_{\rm H} = 187~{\rm cm^{-3}}$.

We enrich the gas cloud with a fixed amount of metals homogeneously distributed over the entire cloud at this moment. We run and examine two cases with metallicity $Z = 10^{-3}~Z_{\odot}$ or $10^{-2}~Z_{\odot}$. \citet{2018MNRAS.475.4378C} show that such a situation is realized when a primordial cloud is engulfed internally by a supernova explosion. 
We also run a reference simulation where the subsequent evolution is followed with the primordial composition. 
Further cloud collapse is followed with the same method as in \citet{2016MNRAS.463.2781C}, using a modified version of GADGET-2 including additional physics that operates with heavy elements. 
We consider an extended chemistry network with 27 gas-phase species, $\rm H^+$, $\rm e$, $\rm H$, $\rm H^-$, $\rm H_2$, $\rm D^+$, $\rm D$, $\rm HD$, ${\rm C^{+}}$, $\rm C$, $\rm CH$, $\rm CH_2$, $\rm CO^{+}$,  $\rm CO$, $\rm CO_2$, $\rm O^+$, $\rm O$, $\rm OH^+$, $\rm OH$, $\rm H_{2}O^+$, $\rm H_{2}O$, $\rm H_{3}O^+$, $\rm O_{2}^{+}$, $\rm O_{2}$, $\rm Si$, $\rm SiO$, and $\rm SiO_2$. 
We also adopt the model of \citet{1994ApJ...421..615P} for dust grains with the same size distribution of \citet{2000ApJ...534..809O}. 
We do not consider the contribution of ice and organic dust grains for simplicity.
These species are expected to remain only in the inner regions of clouds with $Z=10^{-3}~Z_{\odot}$ or $10^{-2}~Z_{\odot}$ because their sublimation temperatures are $\sim 300-700~{\rm K}$ \citep{1994ApJ...421..615P}. In \citet{1994ApJ...421..615P}, 55\% of carbon is locked up into refractory organics. Although the budget of carbonaceous grains is uncertain, we assume that the same fraction of carbon is locked in amorphous carbon grains. The optical constants of amorphous carbon and ice/organics are almost similar \citep{2008ApJ...684.1343N}. Therefore, the thermal evolution of the cloud is almost independent on species of
carbonaceous grains.


The cloud collapses gravitationally in a well-known, self-similar fashion \citep{1969MNRAS.145..271L, 1998ApJ...508..141O}.
We do not observe fragmentation in any of our simulations before the central density reaches $10^{10}-10^{11}~{\rm cm^{-3}}$. Then the minimum local Jeans length becomes $\simeq 30$~au, which is comparable to the size of the sink cell we put afterward. At this point, we switch to 2D RHD simulations to follow the evolution in the accretion phase. We remap the 3D data of the final snapshot to 2D data with the polar coordinate by the following method \citep{2014ApJ...781...60H}. We compute the angular momentum vector of the gas contained in the central 0.01~pc region around the density peak and fix the cloud rotation axis.
We then estimate the physical values in a plane containing the angular momentum vector, and take the average for opposite points with respect to the mid-plane. We finally assign the obtained values at the cell centers. Since our choice of the mapped plane is arbitrary, in Appendix \ref{map_3Dto2D}, we examine how the physical quantities in the original 3D data deviate from  
the distribution of the single plane used in 2D simulations. 
We also investigate the effect of varying the radius to compute the angular momentum vector, which is set to be 0.01~pc as our fiducial choice.


\subsection{Numerical method: protostellar accretion}
\label{initial_conditions2}

We use a modified version of the publicly available (magneto-) hydrodynamics simulation code PLUTO 4.1. Our version incorporates self-gravity, radiative transfer, non-equilibrium chemistry and protostellar evolution, which have been successively implemented in a series of works including \citet{Kuiper2010, Kuiper2012}, \citet{2016ApJ...824..119H}, and \citet{2018A&A...616A.101K}. 
These studies consider massive star formation with $Z = Z_\odot$ and $Z = 0$.


We adopt 2D polar coordinates assuming axial and mid-plane symmetry as in previous studies \citep[e.g.,][]{Kuiper2010}.
As the radial inner boundary, we adopt a sink cell which masks an accreting protostar evolving at the grid center. We assume that the protostar grows in mass according to the amount of the gas flowing into the sink. The radiation emitted by the protostar is injected back into the computational domain, which causes radiative feedback against the accretion flow. We solve the following governing equations of compressible hydrodynamics:
\begin{eqnarray}
	\frac{\partial \rho}{\partial t} + \nabla \cdot \left( \rho \bm{v} \right) = 0, \label{1.1}
\end{eqnarray}
\begin{eqnarray}
	\frac{\partial \left( \rho \bm{v} \right)}{\partial t} + \nabla \cdot \left( \rho \bm{v} \otimes \bm{v} \right) = - \rho \nabla \Phi - \nabla p + \rho \bm{a} , \label{1.2}
\end{eqnarray}
\begin{eqnarray}
	\frac{\partial e}{\partial t} + \nabla \cdot \left( e \bm{v} \right) = - p \nabla \cdot \bm{v} + \Gamma - \Lambda, \label{1.3}
\end{eqnarray}
\begin{eqnarray}
	p = \left( \gamma - 1 \right) e, \label{1.4}
\end{eqnarray}
where $\rho$, $p$, $\bm{v}$, $\Phi$, $e$, $\Gamma$ and $\Lambda$ are the density, pressure, velocity, gravitational potential, the specific energy density, the heating and cooling functions.
In Equation \eqref{1.2}, $\bm{a}$ is the acceleration source term induced by viscosity ($\bm{a_{\rm vis}}$) and by radiation pressure ($\bm{a_{\rm rad}}$).

\subsubsection{Non-equilibrium chemistry and thermal processes} \label{chem_therm}

In addition to Equations \eqref{1.1}-\eqref{1.4}, we solve the non-equilibrium chemistry network with 9 species of H, $\rm H_2$, $\rm H^{-}$, $\rm H^{+}$, e, $\rm C^{+}$, O, $\rm O^{+}$ and $\rm O^{2+}$ to compute the relevant heating and cooling rates in Equation \eqref{1.3} \citep[e.g.,][]{2016ApJ...824..119H,2018ApJ...857...57N, 2018ApJ...865...75N,2019ApJ...883..127N}. 
We use the different chemical network from that of the 3D simulations used for the collapse phase.
In the 2D simulations, we assume that the carbon and oxygen are in the forms of $\rm C^{+}$, O, $\rm O^{+}$ and $\rm O^{2+}$, and we ignore other species considered in the previous simulations.
We also calculate the specific heat $\gamma$ as a function of temperature and chemical abundances following \cite{1998ApJ...508..141O}. Complete lists of the adopted chemistry network and relevant thermal processes are described in Appendix \ref{chemapp}. We here provide a brief summary. 


In a low-metallicity gas cloud, thermal processes by heavy elements significantly 
alter the structure of an accreting envelope around a protostar.
In our 2D RHD simulations, it is infeasible to solve the full set of the chemical network such as that used for the collapse phase and to incorporate all the relevant thermal processes because of high computational cost.
We use a minimal model that reproduces the temperature evolution at the center of a collapsing cloud with various metallicities \citep[e.g.,][]{2005ApJ...626..627O}, and thus the difference in these chemical models does not significantly change our results.  
Specifically, we incorporate the radiative cooling via [O{\sc i}] $63~\micron$ and [C{\sc ii}] $158~\micron$ fine-structure line emission and dust thermal emission. 
We assume that the dust-to-gas mass ratio is 0.01 at $Z=Z_{\odot}$ and it decreases in proportion to the metallicity for $Z < Z_\odot$ for simplicity, though some theoretical studies suggest that the dust-to-gas mass ratio does not linearly scale with metallicity due to slow dust growth \citep{2013MNRAS.432..637A, 2014A&A...563A..31R}. 
We also consider the cooling via O{\sc ii} and O{\sc iii} line emission, which predominantly operate in H{\sc ii} regions at $Z \sim Z_{\odot}$ \citep[e.g.,][]{1989agna.book.....O}. 
However, these metal line cooling becomes ineffective at lower metallicity due to their lower abundances, and free-free emission and hydrogen recombination dominate the cooling below a critical metallicity at $Z\sim 10^{-2}Z_{\odot}$ \citep{2011piim.book.....D}. 
Owing to these coolings, temperature of the photoionized gas varies with different metallicities: $T \simeq 8000$~K for $Z = Z_\odot$ and $T \simeq 3 \times 10^4$~K for $Z \lesssim 10^{-2}~Z_\odot$. To obtain the relative abundances among O{\sc i}, O{\sc ii}, and O{\sc iii}, we first assume that the relative abundance of O{\sc i} is the same as that of H{\sc i} because the ionization energies of O{\sc i} and H{\sc i} are almost equal, and the relative O{\sc ii} and O{\sc iii} abundances are then calculated from the balance between the ionization of O{\sc ii} and recombination of O{\sc iii} (also see Appendix~\ref{ssec:oxy}).

\subsubsection{Angular momentum transport via $\alpha$-viscosity}
\label{alpha_vis}

The accreting gas with finite angular momentum does not directly hit the stellar surface, but first lands on a circumstellar disk. 
The central star mostly accretes the gas through the disk, where the angular momentum transport is driven by gravitational torque.
Although the torque is caused by non-axisymmetric mass distribution induced by the gravitational instability, such an effect does not
enter automatically in our 2D simulations assuming axial symmetry. We use the so-called $\alpha$-viscosity model \citep{1973A&A....24..337S} to mimic this effect, adding relevant acceleration and heating source terms in Equations \eqref{1.2} and \eqref{1.3} respectively \citep[e.g.,][]{2011ApJ...732...20K}.
We assume the spatial distribution of the $\alpha$-parameter as
\begin{eqnarray} 
\alpha(R, Z) = \alpha_0 \exp \left( - \frac{Z}{H(R)} \right), 
\label{eq:alpha}
\end{eqnarray}
where $(R,Z) = (r \sin \theta, r \cos \theta)$ and $H(R)$ is the scale height of the accretion disk at $R$. We calculate the scale height with an approximate formula $H(R) \simeq c_s/\Omega$, where $c_s$ is the sound speed and $\Omega$ is the rotational angular velocity on the equator. In Equation \eqref{eq:alpha}, we estimate $\alpha_{0}$ as 
\begin{eqnarray}
\alpha_{0} = \alpha_{\rm max} \exp(- Q^4), 
\label{eq:alpha0}
\end{eqnarray}
where $Q$ is the Toomre $Q$ parameter \citep{1964ApJ...139.1217T} 
\begin{eqnarray}
Q = \frac{c_{\rm s}\Omega}{\pi G \Sigma}, 
\label{eq:q}
\end{eqnarray}
and we set $\alpha_{\rm max} = 2$ in our models \citep{2013ApJ...770...71T,2010ApJ...713.1143Z, 2014ApJ...781...60H}.
With the above modeling, large $\alpha$-viscosity operates only in the gravitationally unstable disk where $Q \lesssim 1$.


A limitation of our method is inability to accurately capture the disk fragmentation, which has been often observed in 3D simulations of the primordial star formation \citep[e.g.,][]{Machida08,Stacy10,Clark11,2016ApJ...824..119H,Susa19} and present-day star formation \citep[e.g.,][]{2017MNRAS.464L..90M, 2018MNRAS.473.3615M} 
though the fragmentation can be suppressed by rapid gas accretion \citep[]{2020ApJ...892L...4L}. 
Fortunately, simulations also show that, even for such cases, most of fragments rapidly migrate inward through the disk and accrete on the central star \citep[e.g.,][]{2012MNRAS.424..399G,Chon19}.  
Thus our simulations are considered as limiting cases where the fragments, if any, eventually all accrete onto the central star.
We have also found the ``ring-like'' fragmentation even in our 2D simulations for $Z = 10^{-3}~Z_\odot$ and $10^{-2}~Z_\odot$.  
It may be a signature of the physical disk fragmentation promoted by additional cooling via dust thermal emission \citep[e.g.,][]{2014MNRAS.439.1884T}, but  is more likely a spurious one because the Jeans length is only poorly resolved for such cases. In order to prevent this, we monitor Toomre $Q$ parameter to set a temperature floor to satisfy $Q \geq Q_{\rm min} = 0.6$. The floor values are represented by the corresponding minimum sound speed, 
\begin{equation}
c_{\rm s,min} (R,Z) = \frac{Q_{\rm min} \pi G \Sigma}{\Omega} 
\exp \left( - \frac{Z}{H(R)} \right) ,
\end{equation}
where we assume the same spatial variation as in Equation \eqref{eq:alpha}. 
We compute the surface density $\Sigma$ at a given radius by vertically integrating the mass distribution over the range of $75^{\circ} \leq \theta \leq 90^{\circ}$.
We note that such a temperature floor has no effect in the primordial cases with $Z = 0$.

\begin{table*}
 	\caption{Models considered}
 	\label{tab1}
 	\centering

\begin{tabular}{|l|c|c|c|c|c|c|} \hline \hline
model & $N_{\rm R} \cdot N_{\theta}$ & $r_{\rm sink} \, [ \,  {\rm au} \, ] $ & $r_{\rm out} \, [ \,  {\rm pc} \, ] $  & $Z_* \, [ \, Z_{\odot} \, ]$ & $M \, [\, M_{\odot} \,]$ &  Feedback \\ \hline 
Prm & $150 \cdot 40$ & 30 & 1 & $0$ & $560$  &  RP${}^\text{a}$, PI${}^\text{b}$ \\
Prm\_NO & $150 \cdot 40$& 30 & 1 & $0$ & $> 820$ &  NO${}^\text{c}$ \\
-3Sol & $150 \cdot 40$& 30 & 1 &  $10^{-3}$   & $440$ & RP, PI \\
-3Sol\_NO & $150 \cdot 40$ & 30 & 1&  $10^{-3}$   &  $ > 900$ & NO \\
-2Sol & $150 \cdot 40$& 30 & 3 & $10^{-2}$   & $190$  & RP, PI \\
-2Sol\_NO & $150 \cdot 40$ & 30 & 3& $10^{-2}$   & $>670$  & NO  \\
-2Sol\_RP & $150 \cdot 40$& 30 & 3 &  $10^{-2}$  & $> 500$ & RP \\
-2Sol\_PI & $150 \cdot 40$& 30 & 3 &  $10^{-2}$  & $240$  & PI \\
-2Sol\_PI2 & $150 \cdot 40$& 30 & 3 &  $10^{-2}$  & $180$ & PI, w/o dust${}^\text{d}$ \\
-1Sol\_from-2Sol & $150 \cdot 40$ & 30 & 3& $10^{-2} \to 0.1$~${}^\text{e}$  &  $290$  & RP, PI \\
Sol\_from-2Sol & $150 \cdot 40$ & 30 & 3& $10^{-2} \to 1$~${}^\text{f}$   &  $360$  & RP, PI \\
Prm\_from-3Sol & $150 \cdot 40$& 30 & 1 & $10^{-3} \to 0$~${}^\text{g}$   &  $400$  & RP, PI \\
Prm\_highres & $240 \cdot 60$& 30 & 1 & $0$  & 520 & RP, PI \\
Prm\_rsink10 & $150 \cdot 40$& 10 & 1 & $0$  & & RP, PI   \\

\hline 
\end{tabular}
     \begin{minipage}{1 \hsize}
      Notes -. Column 2: cell numbers, Column 3: sink radii, Column 4: position of the radial outer boundary, Column 5: metallicity,  Column 6: final stellar masses, Column 7: feedback mechanisms included
      
      ${}^\text{a}$Suffix "RP" represents radiation-pressure feedback
      
      ${}^\text{b}$Suffix "PI" represents photoionization feedback
      
      ${}^\text{c}$Suffix "NO" represents no radiation feedback included in simulations
      
      ${}^\text{d}$Suffix "w/o dust" represents the dust attenuation of ionizing photons is not included
      
      ${}^\text{e-g}$  The metallicity is changed as indicated at the beginning of the accretion stage
      \\
     \end{minipage}
     
\end{table*}


\subsubsection{Protostellar evolution}\label{protostellar_evolution}

A protostar emits a copious amount of radiation once grown massive. 
The strength of the resulting radiative feedback against the accretion flow thus depends on the protostellar evolution. 
We utilize tabulated stellar evolution tracks, which are pre-calculated with constant accretion rates of $10^{-5} - 10^{-1}~M_{\odot}{\rm yr^{-1}}$ at various metallicities, using the numerical model developed in our previous studies \citep{2009ApJ...691..823H,2009ApJ...703.1810H}. The tracks provide the stellar radius $R_*$ and luminosity $L_*$ as functions of the stellar mass and accretion rate. 

The protostellar evolution modeling relying on the pre-calculated tables could show spurious behavior because it ignores accretion histories. In order to alleviate it, we here use the accretion rate averaged over the characteristic evolutionary timescale $\Delta t_*$. We always monitor the so-called Kelvin-Helmholtz (KH) timescale $t_{\rm KH} = (GM_{*}^{2})/(R_{*}L_{*})$ and accretion timescale $t_{\rm acc} = M_* / \dot{M}_*$, and assume
\begin{equation}
 \Delta t_* = 0.1 \times {\rm min} (t_{\rm KH}, t_{\rm acc}).
\end{equation}
At lower metallicity, the radii of main sequence stars are smaller because the inefficiency of the CNO cycle requires the stars to contract more. 
On the other hand, the total luminosity of the protostar with $M_* > 100~{M_{\odot}}$ is close to the Eddington value, which only has a weak metallicity-dependence. Thus the effective temperature of a low-metallicity star is higher, obeying $T_{\rm eff} \propto R_{*}^{-1/2}$, and its emission spectrum is harder than the solar-metallicity counterpart.

\subsubsection{Radiative transfer}
\label{radiation_transfer}

In order to consider both the photoionization and radiation-pressure effects from the accreting protostar, we solve the frequency-dependent radiative transfer for stellar irradiation by the following method. We inject photons from the sink cell to the computational domain at the rate
\begin{equation}
L_{\rm *}^{\rm tot} = L_* + L_{\rm acc},
\end{equation}
where $L_*$ is the stellar luminosity and $L_{\rm acc} \equiv G M_* \dot{M}_* / R_*$ is the accretion luminosity. The radiation spectrum is assumed to be thermal black-body $L_{*, \nu}^{\rm tot} \propto B_\nu (T_{\rm eff})$, where $T_{\rm eff}$ is the effective temperature defined as
\begin{equation}
 T_{\rm eff} = \left( \frac{L_*^{\rm tot}}{4 \pi \sigma R_*^2} \right)^{1/4} .
\end{equation}
Assumption of the thermal black-body radiation could overestimate the emissivity of ionizing photons at $Z \sim Z_\odot$ owing to the line-blanketing effect, especially for stars with $M_* \sim 10~\msun$ \citep[e.g.,][]{Diaz-Miller98}. This approximation should be valid in our cases since we mainly consider the feedback from more massive stars exceeding $100~\msun$. 
As discussed in Section \ref{protostellar_evolution}, the effective temperature is higher for lower metallicity at a given mass: at $M_{*} = 100~M_{\odot}$, for instance, $T_{\rm eff} \simeq 10^{5}~{\rm K}$ for $Z=0$, $T_{\rm eff} \simeq 7 \times 10^{4}~{\rm K}$ for $Z=10^{-2}Z_{\odot}$, and $T_{\rm eff} \simeq 5 \times 10^{4}~{\rm K}$ for $Z=Z_{\odot}$. 
We use a hybrid method where the direct light emitted from the central star and diffuse light re-emitted from the accretion envelope are separately solved \citep[e.g.,][]{2010A&A...511A..81K,Kuiper2010}. We solve the transfer of the stellar direct component by means of the ray-tracing method,
\begin{eqnarray}
    F_{\nu}(r) = \frac{L_{*,\nu}^{\rm tot}}{4 \pi r^2} e^{-\tau_{\nu}}, 
\label{eq:direct_flux}
\end{eqnarray}
where $F_{\nu}(r)$ is the flux at the radial position $r$ and $\tau_{\nu}$ is the optical depth
\begin{eqnarray}
  \tau_{\nu} = 
\int_{r_{\rm in}}^{r} \left[ n_{\rm HI} \sigma_{\rm HI}(\nu) + \rho \kappa_{\rm d} (\nu) \right] dr, 
\label{eq:tau_nu}
\end{eqnarray}
where $\sigma_{\rm HI}(\nu)$ is the H{\sc i} photoionization cross section \citep[e.g.,][]{1989agna.book.....O}, $\kappa_{\rm d} (\nu)$ the dust opacity given by \citet{1993ApJ...402..441L}, and $r_{\rm in}$ the sink radius. 
We assume that the dust-to-gas mass ratio decreases with decreasing metallicities, linearly scaling with $Z$. The dust opacity is determined by the scaled dust-to-gas mass ratio.
 We also assume that photons freely travel without attenuation for $R_* < r < r_{\rm in}$. 
We set 200 logarithmically spaced wavelength bins between $0.03 ~{\rm cm}$ and $1.5~{\rm nm}$, achieving the higher resolution for the shorter wavelengths.
The hydrogen photoionization rate is calculated by integrating the contributions by photons below the Lyman limit, i.e., $91.2~{\rm nm}$ (Appendix~\ref{photoheat}). 
We do not include the diffuse ionizing photons, as their contribution to the disk photoevaporation is negligible compared with the stellar direct component in the massive star formation \citep[e.g.,][]{2011Sci...334.1250H, 2013ApJ...773..155T}.
As in \citet{2016ApJ...824..119H}, we do not incorporate helium ionization for simplicity. We assume that helium is in the atomic state everywhere including H {\sc ii} regions. We consider the transfer of He ionizing photons above 54.4~eV emitted from the protostar, but they are consumed to photoionize hydrogen in our simulations.
Regarding the photodissociation of hydrogen molecules, we do not use $F_{\nu}(r)$ because our wavelength grids are too sparse to resolve the Lyman-Werner bands. We instead consider another monochromatic direct component representing FUV ($11.2~{\rm eV} \leq h \nu \leq 13.6~{\rm eV}$) photons, 
\begin{equation}
 F_{\rm FUV} (r) = \frac{L_{\rm FUV}}{4 \pi r^2} \exp (- \tau_{\rm d, FUV}),
\end{equation}
where $L_{\rm FUV}$ is the luminosity in the FUV wavelength range, and $\tau_{\rm d, FUV}$ is the dust opacity represented by the value at $\lambda = 100~{\rm nm}$. 
The photodissociation rate of hydrogen molecules is estimated with the flux $F_{\rm FUV}$, including the $\rm H_2$ self-shielding.
The details of the calculation are shown in Appendix~\ref{photoheat}


The dust thermal radiation emitted from the accretion envelope is only the diffuse component considered here. We approximately solve the transfer of such diffuse light using the gray flux-limited diffusion (FLD) method \citep{Levermore81}, with the solver developed by \citet{2010A&A...511A..81K}. This computation is performed except for the primordial cases where there are no dust grains. The obtained radiation fields, including both the direct and diffuse components, are used to calculate the acceleration source term $\bm{a}_{\rm rad}$ in Equation \eqref{1.2} and dust temperature $T_{\rm gr}$ (see Appendix~\ref{tdust}).

\subsection{Cases considered}
\label{cases_considerd}

Table \ref{tab1} summarizes the models we consider in the present paper. The fiducial models are Prm, -3Sol, and -2Sol, where the protostellar radiative feedback is fully incorporated for the cases with $Z = 0$, $10^{-3}~Z_{\odot}$ and $10^{-2}~Z_{\odot}$. For comparison, we also run simulations without the feedback at each metallicity. We denote these runs as Prm\_NO, -3Sol\_NO, and -2Sol\_NO. To see the impact of the photoionization and radiation-pressure feedback separately, we also study cases with excluding either of the effects with $Z = 10^{-2}Z_{\odot}$ (-2Sol\_PI and -2Sol\_RP). 
To examine the role of the dust attenuation of ionizing photons, we perform an additional simulation -2Sol\_PI2. 
The cloud properties such as the rotational velocity at the beginning of 2D calculation is an outcome, as a result of its evolution during the early collapse phase, and thus they differ in runs with different values of metallicity. 
To isolate the metallicity effect on radiative feedback, we also perform additional experiments where we start 2D calculation from the same accretion envelope structure. 
In practice, we vary the metallicity after the collapse stage. For instance, Prm\_from-3Sol is the case with $Z=0$ in the 2D calculation but from the envelope structure resulting from the cloud collapse with $Z = 10^{-3}~Z_\odot$.  -1Sol\_from-2Sol and Sol\_from-2Sol are similar cases with $0.1~Z_\odot$ and $1 Z_\odot$, but from the $Z = 10^{-2}~Z_\odot$ envelope structure. 
The accretion envelope for cases -1Sol\_from-2Sol and Sol\_from-2Sol have higher densities than that expected from the thermal evolution during the collapse stage with $Z = 0.1~Z_{\odot}$ and $1~Z_{\odot}$. 
The present-day high-mass star formation supposes a similar situation, where the accretion envelope is much denser than that expected purely by the typical temperature of $\sim 10$~K \citep[e.g.,][]{2003ApJ...585..850M}. The resulting accretion rates are much higher than that expected with $\sim 10$~K, $\sim 10^{-5}~~M_{\odot} {\rm yr^{-1}}$. -1Sol\_from-2Sol and Sol\_from-2Sol correspond to such cases with $0.1~Z_\odot$ and $Z_\odot$, respectively.


In our simulations, the grid cells are logarithmically spaced in the radial direction to improve the spatial resolution near the center, while they are homogeneously distributed in the polar directions. The numbers of grid cells are 150 and 40 for the radial and polar directions for most of the cases. We set the radial outer boundary so that the computational domain contains the gas exceeding $10^3~M_{\odot}$. The radial inner boundary is located at $r_{\rm min} = 30$~au. 
\citet{Kuiper2010} and \citet{Krumholz18} show that, to accurately follow the radiation-pressure effect against the accretion flow, it is necessary to spatially resolve the dust destruction front created around a luminous source. 
The position of the dust destruction front in our simulations is analytically derived as 
\begin{eqnarray}
	R_{\rm d} \simeq 110 ~{\rm au} \left( \frac{L_*^{\rm tot}}{10^{5} ~L_{\odot}} \right)^{1/2},
 \label{eq:rd}
\end{eqnarray}
where $L_*^{\rm tot}$ is the stellar total luminosity, which exceeds $10^{5} ~L_{\odot}$ for $M_* \gtrsim 30~\msun$ \citep{2018MNRAS.473.4754F}.

This means that the dust destruction front is located far outside of $r_{\rm min} = 30$~au, and is well resolved in our simulations. In Appendix \ref{resol}, we present the convergence tests with varying the spatial resolution and the radius of the inner boundary, showing that the results do not significantly change.

\section{Results} 
\label{Results}


\subsection{Early phase: cloud collapse}\label{Cloud_Collapse}

 \begin{figure}
 \begin{center}
 \includegraphics[width=\columnwidth]{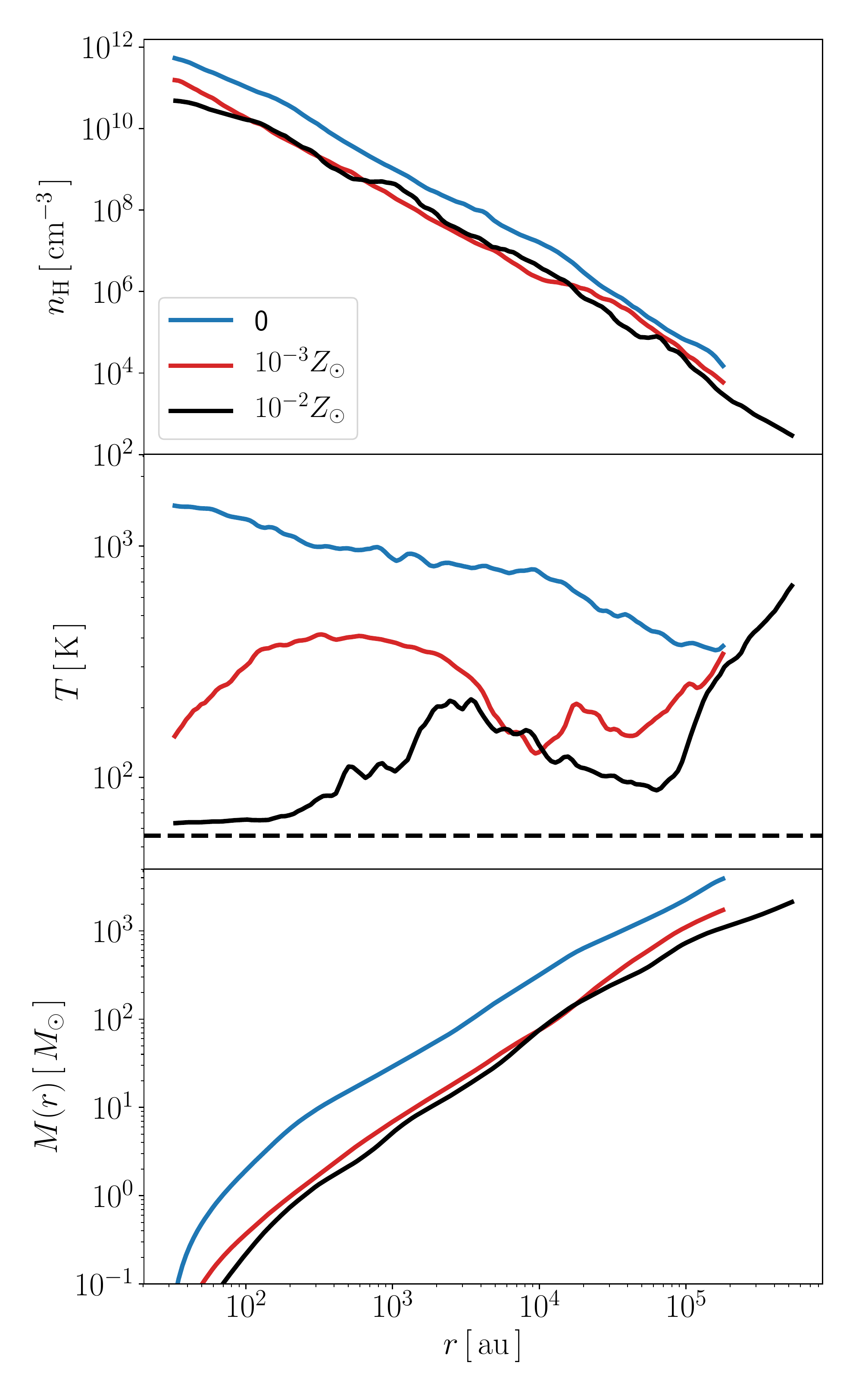}
 \end{center}
 \caption{Radial structure of the accretion envelope at the end of the collapse phase for three different metallicities $Z=0$ (blue), $10^{-3}Z_{\odot}$ (red), and $10^{-2}Z_{\odot}$ (black). Top, middle, and bottom panels present the radial profiles of the number density $n_{\rm H}$, temperature $T$, and the enclosed mass $M(r)$ as functions of the radial distance from the cloud center. The horizontal dashed line in the middle panel represents the CMB temperature at $z = 20.46$.
}
 \label{zu_init2}
 \end{figure}
 \begin{figure}
 \begin{center}
 \includegraphics[width=\columnwidth]{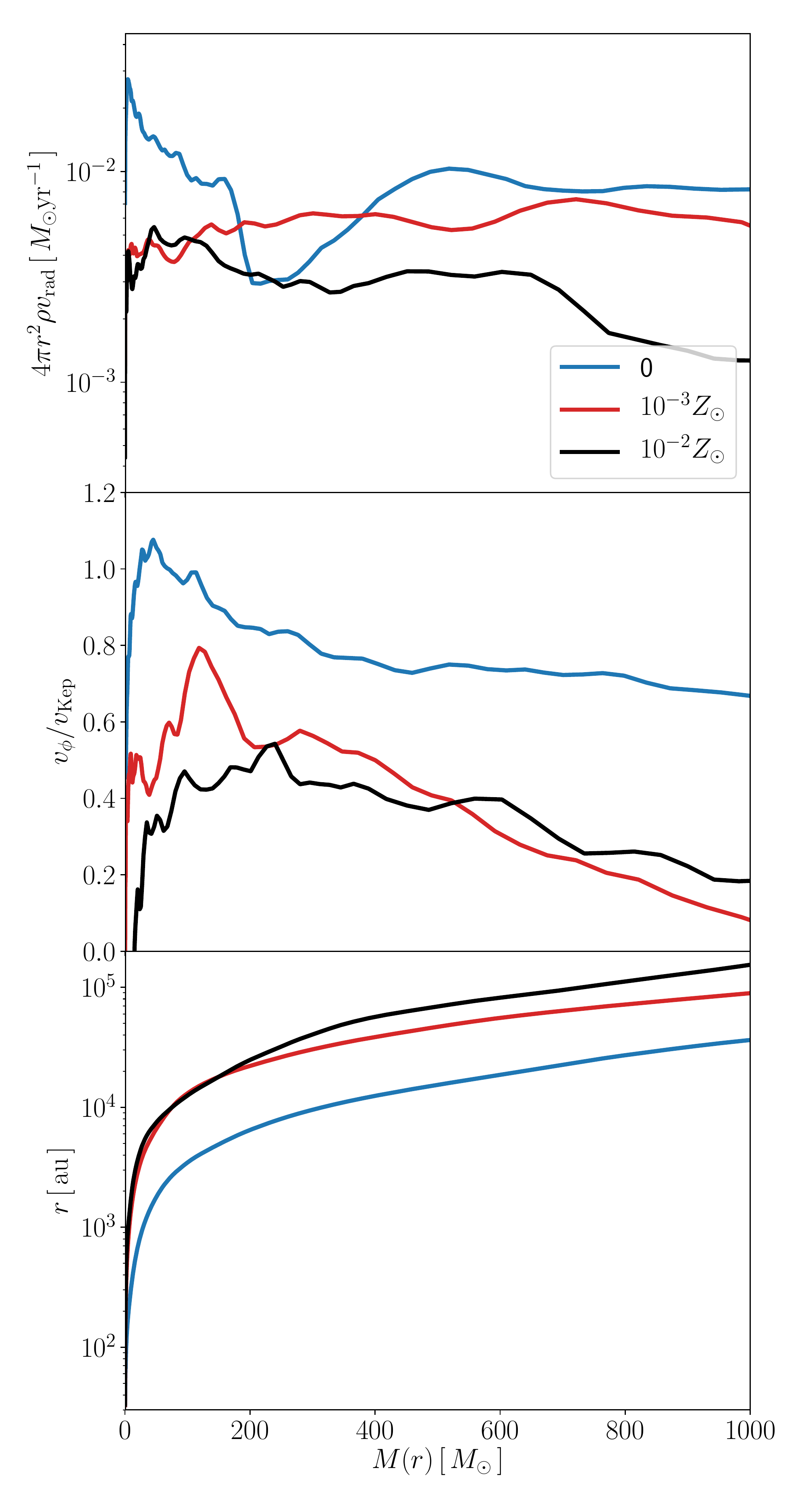}
 \end{center}
 \caption{ Properties of the accretion envelope at the end of the collapse phase for the cases shown in  Fig.~\ref{zu_init2}. The top, middle and bottom panels show the mass inflow rate calculated from the envelope density and velocity structure (also see text), mass-averaged rotational velocity normalized by the Keplerian value  $v_{\rm K} = \sqrt{GM(r)/r}$ and the radial distance from the cloud center as functions of the enclosed mass $M(r)$.
 The colors represent the same cases as in Fig.~\ref{zu_init2}.  }
 \label{zu_init1}
 \end{figure}
 \begin{figure}
 \begin{center}
 \includegraphics[width=7.5cm]{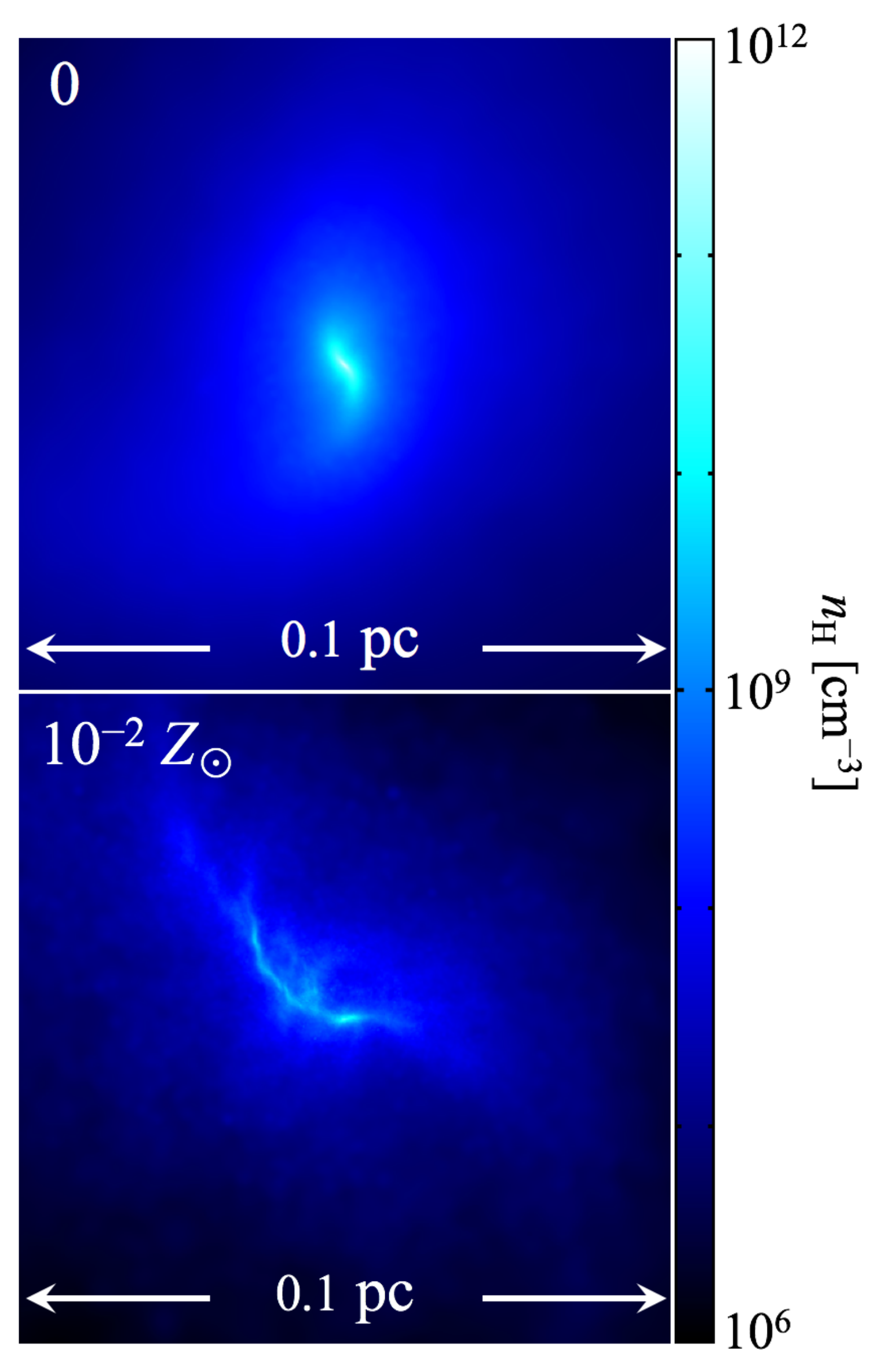}
 \end{center}
 \caption{The cloud morphology at the end of the collapse for the cases with $Z=0$ (top) and $10^{-2}~Z_{\odot}$ (bottom). Each panel show the distribution of the number density on the "face-on" view, i.e., projected on a surface perpendicular to the angular momentum vector. The density peak of the cloud is located at the center of each panel.  
 }
 \label{snap_shot3D}
 \end{figure}

In this section, we describe the evolution in the early collapse stage in the 3D simulations, and variation among the cases with different metallicities $Z=0$, $10^{-3}~Z_\odot$, and $10^{-2}~Z_\odot$.
Figure \ref{zu_init2} shows the radial distributions of the mass-averaged number density and temperature in the accretion envelope at the end of the collapse. 
The temperature distribution at this epoch reflects the thermal evolution during the collapse \citep[e.g.,][]{2005ApJ...626..627O}. For the case with $Z=0$, where the gas cools only via the H$_2$ rovibrational line emission, the temperature gradually rises with increasing density from $T \simeq 400~{\rm K}$ at $n_{\rm H} \sim 10^{4}~{\rm cm^{-3}}$. 
For the cases with $Z = 10^{-3}~Z_{\odot}$ and $10^{-2}~Z_{\odot}$, the temperature decreases with the density because of additional metal cooling. At $Z = 10^{-2}~Z_\odot$, for instance, the temperature drops to $T \sim 100~{\rm K}$ owing to O{\sc i} and C{\sc ii} fine-structure line cooling for $n \sim 10^{3} - 10^{6}~{\rm cm^{-3}}$. The dust cooling becomes effective for $n_{\rm H} \gtrsim 10^{8}~{\rm cm^{-3}}$ or $r \lesssim 10^3$~au, where the temperature approaches the CMB value at $z \simeq 20$.
The top and bottom panels of Figure \ref{zu_init2} show that the density and thus the enclosed mass at a given radius are larger for $Z=0$, but they only slightly differ among the cases with $Z=10^{-3}Z_{\odot}$ and $10^{-2}Z_{\odot}$.
The density distribution in the accretion envelope depends on the thermal evolution of the cloud. Since the gravity and thermal pressure gradient nearly balance each other in the core part during the collapse, a thermal state $(\rho, T)$ is determined such that the distance from the center $r$ becomes equal to the local Jeans length $r \sim \lambda_{\rm J} \propto T^{1/2} \rho^{-1/2}$. 
The density thus distributes as $\rho \propto T r^{-2}$, which is higher with the higher temperature, i.e., the lower metallicity.


Figure \ref{zu_init1} shows the mass inflow rate and average rotational velocity relative to the Keplerian value in the envelope at the end of the collapse. 
We see that the accretion rate is lower for higher metallicities in general. This can be understood with the well-known scaling law $\dot{M}_* \propto T^{3/2}$ (see also Sec.~\ref{introduction}) since the temperature is lower at higher metallicity in the 
envelope (Figure \ref{zu_init2}). This does not, however, explain all the result:
the temperature differing by a factor of $\simeq 5$ between the cases with $Z= 10^{-2}~Z_{\odot}$ and $Z=0$ at $M (r)=$ a few $\times 100~M_\odot$
, the scaling law $\dot{M}_* \propto T^{3/2}$ predicts a roughly 10 times difference in the accretion rates, while the difference in the inflow rates in Figure \ref{zu_init1} is much smaller.
This is because the temperature decreases inward in the $Z= 10^{-2}~Z_{\odot}$ case. 
Consequently, the outward pressure gradient force is weaker, and 
the density and inward radial velocity are higher than in the isothermal case assumed in deriving the scaling relation. 


As seen in the bottom panel of Figure \ref{zu_init1},
rotational velocities are systematically lower for the cases with $Z= 10^{-2}~Z_{\odot}$ and $10^{-3}~Z_{\odot}$. 
This is due to more efficient angular momentum transport during 
the collapse in those cases. 
Figure~\ref{snap_shot3D} shows the cloud morphology for $Z=0$ and 
$10^{-2}Z_{\odot}$ cases.
Whereas the cloud is almost spherical over $\sim 0.1$~pc for $Z = 0$, it becomes more filamentary during the collapse owing to more efficient cooling for $Z = 10^{-2}~Z_{\odot}$ and such 
non-axisymmetric structure causes the gravitational torque 
extracting angular momentum. 

Note that for the case with $Z=0$ the rotational velocity slightly surpasses the Keplerian value at $M(r) < 100~\msun$. The exact profile of $v_\phi/v_{\rm Kep}$ actually depends on the rotation axis for the gas enclosed within a given radius, currently 0.01~pc,
as this axis determines the rotational velocity in 3D data that to be remapped to the 2D data (see Sec.~\ref{initial_conditions}.
We have confirmed that if we take the larger radius 0.1~pc to fix the rotational axis the $v_\phi$ profile is slightly modified and super-Keplerian velocities disappear.  In any case, however, $v_\phi$ is nearly Keplerian at maximum and results are hardly affected by this uncertainty. 
In Appendix \ref{map_3Dto2D}, we further describe the difference 
when a larger radius 0.1~pc is used for fixing the rotational axis.

\subsection{Late phase: protostellar accretion}\label{prostellar_accretion}
Next, we present the 2D simulation results of the subsequent protostellar accretion stage at each metallicity.
We first describe the models where both the photoionization and radiation-pressure effects are included with metallicities 
$Z=0, 10^{-3}~Z_{\odot}$, and $10^{-2}~Z_{\odot}$ (i.e., models Prm, -3Sol, and -2Sol, see Table 1).
We, in particular, focus on the case with $Z = 10^{-2}~Z_\odot$, clarifying radiative feedback effects that finally halt the accretion onto the star.


In Figure \ref{zu2.1} we present the mass accretion histories in those models, in comparison with the reference cases where the feedback effects are turned off by hand (models Prm\_NO, -3Sol\_NO, and -2Sol\_NO).
Without the feedback, the stars continue growing steadily by accretion even at the end of the simulation ($600~M_{\odot}$), when  
the gas has not yet depleted in the accretion envelope. 
With radiative feedback, in contrast, the mass accretion is terminated by radiative feedback.
We continue the runs until the accretion onto the protostar almost ceases with
the accretion rate being $\dot{M}_* < 10^{-4}~M_{\odot} {\rm yr^{-1}}$. The final stellar masses are $M_* \simeq 560~M_{\odot}$, $450~M_{\odot}$ and $190~M_{\odot}$ at $Z=0$, $10^{-3}~Z_{\odot}$ and $10^{-2}~Z_{\odot}$, respectively.
In the final stage, the mass accretion rate gradually decreases in all these cases.
We fit the accretion histories after $\dot M $ becomes less than $2 \times 10^{-4}~M_{\odot}{\rm yr^{-1}}$ with linear functions of the elapsed time.
The actual values are $3.8$, $7.5$, and  $6.1 \times 10^{-4}~M_{\odot}{\rm yr^{-1}}{\rm Myr^{-1}}$ at the metallicities of $Z=0$, $10^{-3}$, and $10^{-2}~Z_\odot$.
In all the cases, the stellar mass increases only by $<15~M_{\odot}$ until the accretion rate decreases to 0. Thus we ignore the remaining component of the envelope after the accretion rate drops below $10^{-4}~M_{\odot}{\rm yr^{-1}}$.
In what follows, we investigate metallicity dependence of radiative feedback and resulting final stellar masses in more detail. 

\subsubsection{Accretion flows in cases without protostellar feedback}
\label{accretion_to_disk}

 \begin{figure}
 \begin{center}
 \includegraphics[width=\columnwidth]{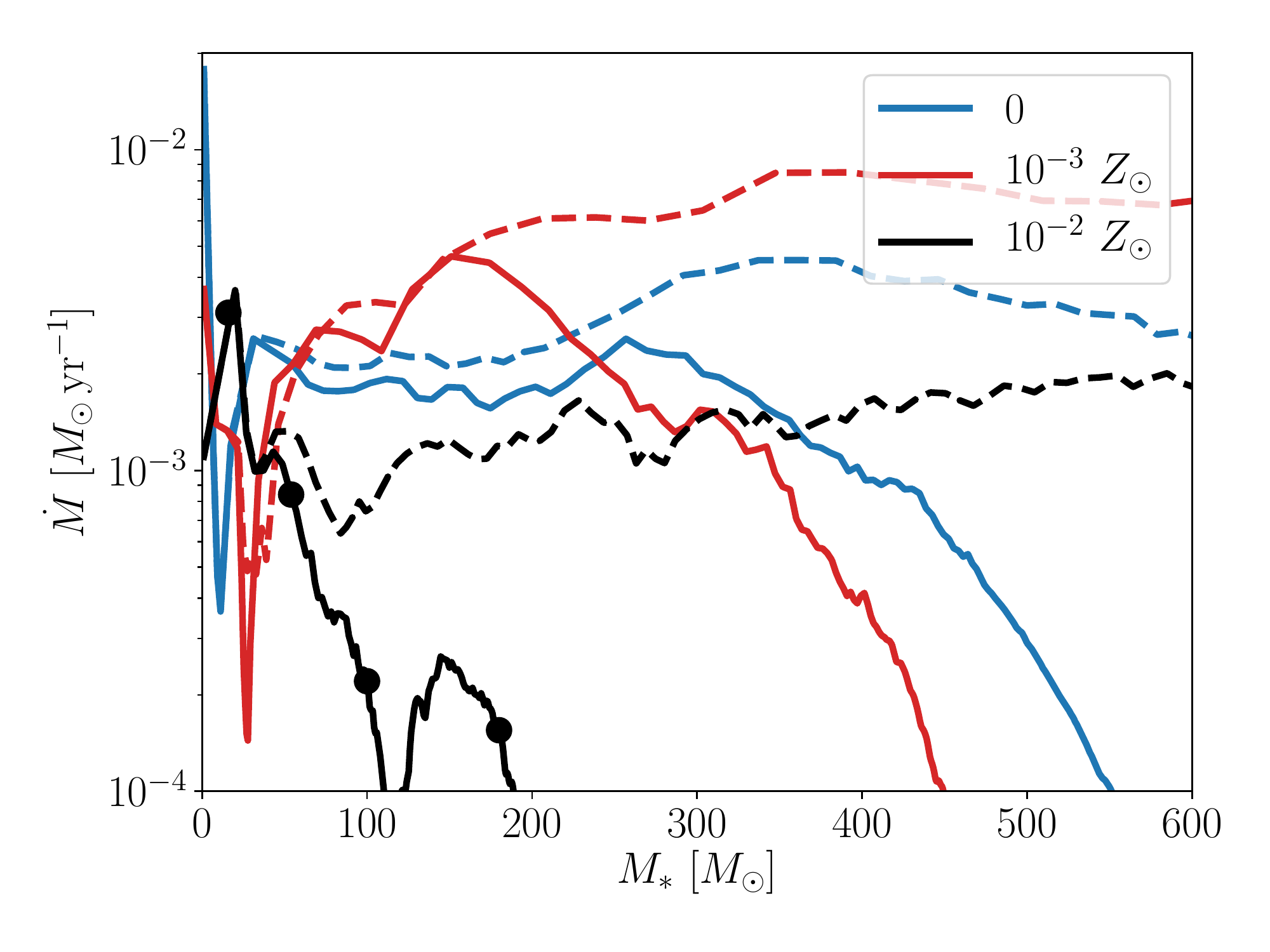}
 \end{center}
 \caption{The mass accretion histories in the protostellar accretion phase for the cases with metallicities $Z=0$, $10^{-3}Z_{\odot}$, and $10^{-2}Z_{\odot}$. 
 Both the cases with and without radiative feedback are shown by solid (for models Prm, -3Sol, and -2Sol) and dashed lines (for Prm\_NO, -3Sol\_NO, and -2Sol\_NO), respectively. 
 The filled circles on the line for the case -2Sol represent the four epochs when the snapshots are presented in Figure~\ref{zu1}.
 }
 \label{zu2.1}
 \end{figure}
 \begin{figure}
 \begin{center}
 \includegraphics[width=\columnwidth]{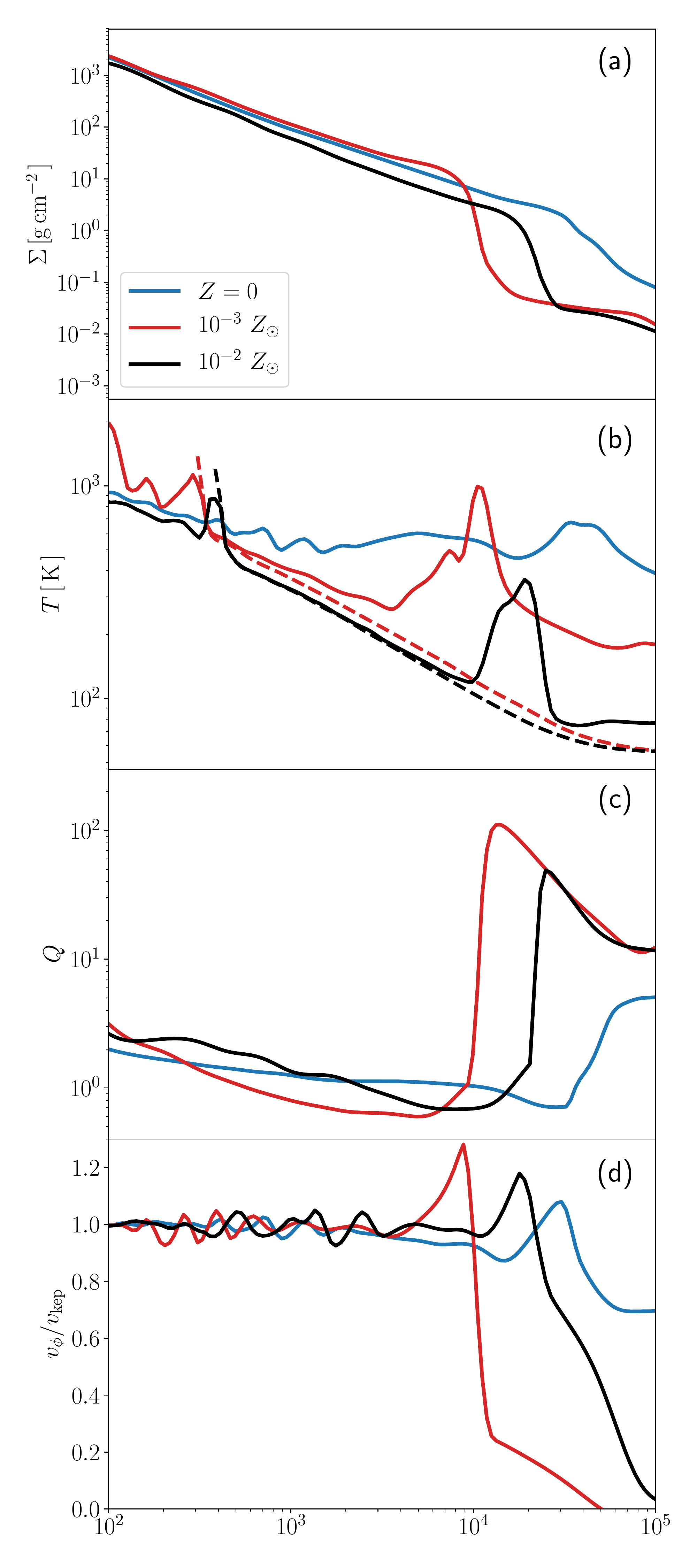}
 \end{center}
 \caption{
The radial structure of the accretion disk when the stellar mass is $\simeq 150~M_{\odot}$ for different metallicities. 
The panels (a) and (b) show the radial structure of the surface density and temperature respectively.
In the panel (b), the solid and dashed lines represent the gas and dust grains temperature. 
 The panels (c) and (d) present Toomre Q parameter and the ratio of the azimuthal velocity relative to the Keplerian velocity against the radial distance from the central star. 
 In each panel the blue, red, and black lines represent the cases with $Z=0$, $10^{-3}~Z_{\odot}$, and $10^{-2}~Z_{\odot}$.
 }
 \label{zu3.2.1}
 \end{figure}

Before examining the feedback effect, 
we first study the cases without the feedback (i.e., Prm\_NO, -3Sol\_NO, and -2Sol\_NO) to see the typical accretion rate at each metallicity. 
Figure \ref{zu2.1} shows that the rate for $Z=0$ (case Prm\_NO) is somewhat higher than that for $10^{-3}~Z_\odot$ (-3Sol\_NO), 
whereas the opposite trend is expected from 
the envelope structure at the end of the collapse (top panel of Figure \ref{zu_init1}). 
This discrepancy comes from the difference in rotation velocities $v_\phi$ (bottom panel of Figure \ref{zu_init1}) as higher $v_\phi$ requires more efficient angular momentum transfer through the circumstellar disk for the matter to accrete onto the protostar.


Indeed, formation of a large disk and its growth around the central protostar is observed in all the simulation runs. Figure \ref{zu3.2.1} shows the disk structure at the stellar mass of $M_* = 150~M_{\odot}$. In all the cases, the rotation is almost Keplerian inside $r \lesssim 10^{4}- 3 \times 10^{4} ~{\rm au}$, which corresponds to the disk outer edge. 
At this radius, the surface density $\Sigma$ also sharply rises and Toomre $Q$ parameter drops accordingly. 
Although the $Q$ value is nearly constant at the order of unity within the disk in all the cases, 
it is higher in the $Z=0$ case by a factor of a few than in the other cases at $10^3~{\rm au} \lesssim r \lesssim 10^4~{\rm au}$. 
This results in less efficient angular momentum transport by gravitational torque in 
the $Z=0$ case from equations \eqref{eq:alpha} - \eqref{eq:q} and thus the flows in the disk has the higher rotation velocities $v_{\phi}$.


The radial profiles of the gas and dust temperatures in the disk mid-plane are presented in the bottom panel of Figure \ref{zu3.2.1}. 
Generally, the gas temperature is lower for higher metallicity, particularly at the outer region, owing to the difference in the dominant cooling agent.
In the $Z=0$ case,  $\rm H_{2}$ line cooling balances with the viscous heating and keeps the temperature at 
$T \simeq 600~{\rm K}$ throughout the disk, while the dust cooling is dominant for $Z=10^{-3}Z_{\odot}$ and $10^{-2}Z_{\odot}$. 
At the disk outer edge, the gas is heated by the accretion shock. 
Then, at high enough densities, the gas and dust 
are thermally coupled by frequent collisions, and the gas temperature is lowered to the dust temperature (see Appendix~\ref{tdust}). 
Note that the dust destruction front is located at $r \simeq 300~{\rm au}$, where the dust grains are sublimated. Inside the dust destruction front, the gas temperatures are similar among the cases with different metallicities due to the lack of dust cooling. 
  
\subsubsection{Radiative feedback effect on the protostellar accretion}
\label{radiative_feedback}

 \begin{figure*}
 \begin{center}
 \includegraphics[width=140mm]{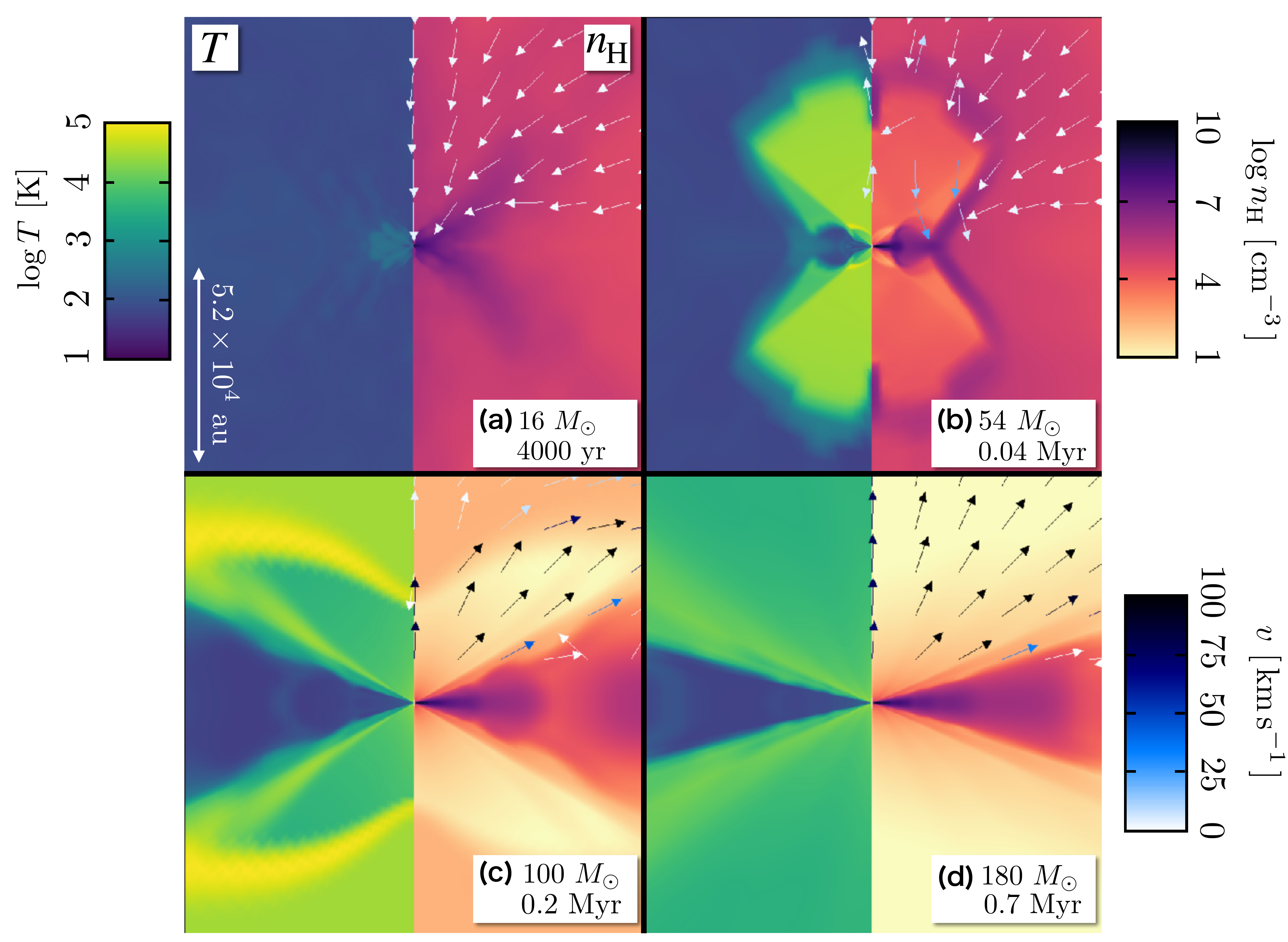}
 \end{center}
 \caption{
Time evolution of the accretion envelope structure around the protostar at $Z = 10^{-2}~Z_\odot$ (model -2Sol). The four panels show the snapshots at the protostellar mass (a) $M_* \simeq 16~M_{\odot}$, (b) $54~M_{\odot}$, (c) $82 ~M_{\odot}$, and (d) $140~M_{\odot}$. The time since the birth of the protostar is also indicated in the bottom-right corner. The distributions of the number density $n_{\rm H}$ (right) and gas temperature $T$ (left) are presented. The arrows shown in the upper-right part represent the velocity vectors.
}
 \label{zu1}
 \end{figure*}
 \begin{figure}
 \begin{center}
 \includegraphics[width=\columnwidth]{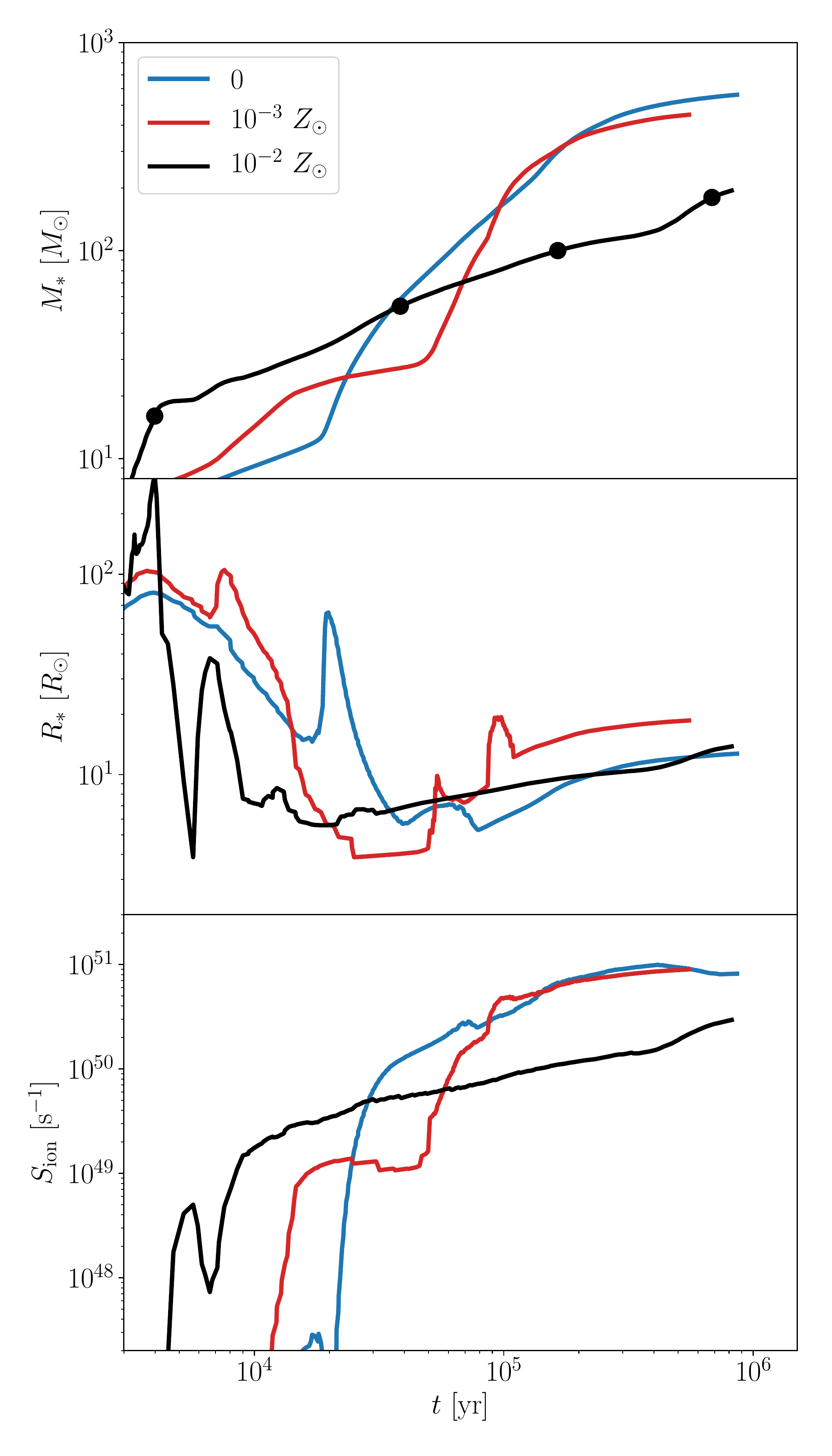}
 \end{center}
 \caption{Protostellar evolution for different metallicities $Z=0$ (model Prm, blue line), $10^{-3}~Z_{\odot}$ (-3Sol, orange), and $10^{-2}~Z_{\odot}$ (-2Sol, black). Plotted are the stellar mass (top panel), radius (middle), and ionizing photon production rate (bottom) as a function of the elapsed time since the birth of the protostar. In the top panel, the filled circles on the black curve represent the four epochs for which the snapshots are presented in Figure~\ref{zu1}.
}
 \label{zu_stellarevo}
 \end{figure}
 \begin{figure}
 \begin{center}
 \includegraphics[width=\columnwidth]{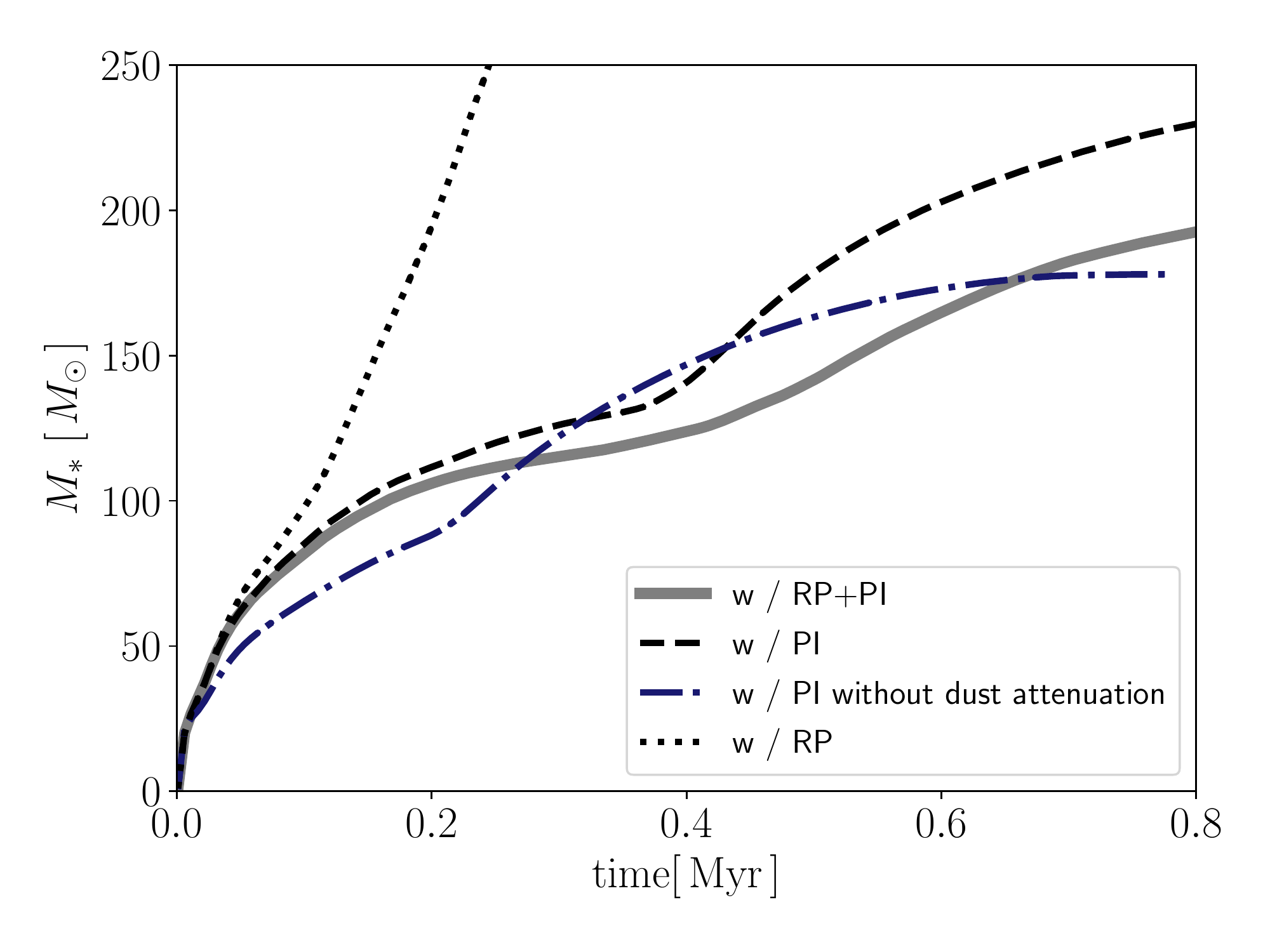}
 \end{center}
 \caption{
 Impact of radiative feedback by photoionization and by radiation pressure on the stellar mass growth for $Z = 10^{-2}Z_{\odot}$. The gray solid line represents 
 the case where both those effects are taken into account (-2Sol). The other lines are for the reference cases where one of the feedback mechanisms is omitted; 
 only with the radiation-pressure effect (black dotted, -2Sol\_RP), 
 only with the photoionization effect (dark blue dot-dashed, -2Sol\_PI), 
 and only with the photoionization effect without the dust attenuation of ionizing photons (dashed line, -2Sol\_PI2). 
 }
 \label{zu3.2}
 \end{figure}

Here we see the protostellar radiative feedback
effect on the accretion flows. 
Figure~\ref{zu1} shows the structure of the gas around the protostar in the case of $Z=10^{-2}~Z_{\odot}$ (model -2Sol).
As seen in Figure \ref{zu1}-a, the protostar accretes the gas through the circumstellar disk. The stellar mass rapidly increases at a high accretion rate $>10^{-3}~M_\odot~{\rm yr^{-1}}$ in this early stage (also see Fig.~\ref{zu2.1}). By the epoch when the stellar mass reaches $\sim 50~M_\odot$, the ionizing-photon emissivity becomes high enough to create bipolar H{\sc ii} regions or cavities (Fig. \ref{zu1}-b). 
The temperature inside the cavities 
exceeds $10^{4} ~{\rm K}$ due to the photoionization heating.  
Dynamical expansion of the H{\sc ii} regions blows away the gas in the accretion envelope. 
The mass supply from the envelope to the accretion disk is also interrupted for a while.
As a result, the accretion rate is reduced from the initial value of $3 \times 10^{-3} M_{\odot} {\rm yr^{-1}}$ by an order of magnitude by the time the stellar mass reaches $50 M_{\odot}$ (Figure~\ref{zu2.1}).

Within the bipolar H{\sc ii} region (Figure~\ref{zu1}-c), we can see a secondary bubble expanding, in which the gas is escaping outward at high velocity $\sim 100~{\rm km~s^{-1}}$ and the density is very low at $\sim 10~{\rm cm^{-3}}$. 
The expansion of this secondary bubble is driven by the radiation pressure exerted on dust grains. Since the grains are dynamically well coupled with the gas, the gas is also efficiently accelerated outward.
As a result of the expansion of the secondary bubble, the temperature of the photoionized gas is lowered to several $\times~10^3$~K by adiabatic cooling.


Note that, in the primordial star formation, the accretion rate is reduced by the expansion of the \hii regions and no secondary bubble is observed. 
Although the radiation-pressure feedback has been known to be effective in the present-day massive star formation, a detailed mechanism differs from our case: 
while the radiation forces both by the stellar direct light and by the diffuse radiation re-emitted by the dust are equally important in present-day star formation, the latter is negligible in our case. 
The radiation force by direct light can be written relative to the gravity as
\begin{align}
	\frac{F_{\rm rad}}{F_{\rm grav}} = 5.3 \left( \frac{\kappa_{\rm UV, \odot}}{400 ~{\rm cm^2 g^{-1}}} \right)  \left( \frac{L_*}{8.0 \times 10^6 ~L_{\odot}} \right) \left( \frac{M_*}{350 ~ M_{\odot}} \right)^{-1} \left( \frac{Z}{10^{-2}~Z_{\odot}} \right), \label{1114.1}
\end{align} 
where $\kappa_{\rm UV, \odot}$ is the dust opacity for UV photons at $Z = Z_\odot$, 
and that by the diffuse radiation is
\begin{align}
	\frac{F_{\rm rad}}{F_{\rm grav}} = 0.04 \left( \frac{\kappa_{\rm IR, \odot}}{3 ~{\rm cm^2 g^{-1}}} \right)  \left( \frac{L_*}{8.0 \times 10^6 ~L_{\odot}} \right) \left( \frac{M_*}{350 ~ M_{\odot}} \right)^{-1} \left( \frac{Z}{10^{-2}~Z_{\odot}} \right), 
\label{190903.1}
\end{align} 
where $\kappa_{\rm IR, \odot}$ is the dust opacity for infrared photons at $Z = Z_\odot$, for which we assume the dust temperature of $300$~K.
In fact, Figure \ref{zu1} shows that only the gas directly illuminated by the stellar light is accelerated along the poles.

Whereas the bubble expansion blows away the gas in the envelope in polar directions, in a shade of the disk the flow is protected from the direct star light and the accretion still continues. Indeed, Figure~\ref{zu2.1} shows that the accretion rate drops just after the stellar mass exceeds $100~\msun$, but it returns to the level of $\sim 10^{-4}~M_\odot~{\rm yr^{-1}}$. Although the exact behavior of the accretion history should depend on the prescription of the viscosity (Sec.~\ref{alpha_vis}), the stellar mass further increases at such low rates. 
 Eventually, however, even the flows near the equator are disturbed and reversed (Figure~\ref{zu1}-d). The mass supply from the envelope to the disk is first shut off, and then the protostar only swallows the gas retained in the disk. This slow accretion continues for several $\times 10^5$ years at a gradually decreasing rate. The final stellar mass is $\simeq 190~M_{\odot}$ in this case.

The protostellar evolution for the same model (-2Sol) is presented in 
Figure \ref{zu_stellarevo} (black line). The stellar radius is initially very large $>100~R_\odot$ but soon decreases within the initial $10^4$ years (top panel) as a result of Kelvin-Helmholtz (KH) contraction, which continues until the stationary nuclear burning begins, i.e., the star reaches the main sequence, at $M_* \simeq $ a few $\times~10~M_\odot$ \citep[e.g.,][]{2003ApJ...589..677O}. 
The emissivity of ionizing photons increases dramatically during the KH contraction (bottom panel), which is accompanied by the appearance of the bipolar H{\sc ii} regions at $t \simeq$ a few $\times~10^4$ years (Fig. \ref{zu1}-b). 
Note that since the final stellar mass is $\simeq 190~\msun$, the star acquires most of its mass after reaching the main-sequence star.
The ionizing photon emissivity rises gradually with increasing stellar mass until the accretion is finally shut off.


We have seen that the photoionization and radiation-pressure effects, such as the expansion of the \hii regions and the secondary bubble, have dynamical effects on the accretion flow. Here we investigate which one is dominant in setting the final stellar mass. For this purpose, we present the stellar mass growth histories in cases with only one of those effects considered in Figure \ref{zu3.2}. With both the photoionization and radiation-pressure effects, the stellar mass reaches $190M_{\odot}$ in 0.8 Myr as seen above (-2Sol, solid). 
With only the radiation-pressure feedback taken into account (-2Sol\_RP, dotted), the stellar mass rapidly increases and exceeds $250~\msun$ in $\simeq 0.2$~Myr.
In contrast, with the photoionization effect only (-2Sol\_PI, black dashed), 
the final stellar mass hardly changes with a small increment to 
$\simeq 230~\msun$ at $t \simeq 0.8$~Myr.   
The above results clearly demonstrate that the photoionization is the dominant effect in terminating the mass accretion while the radiation pressure plays only a secondary role.
In addition to being a receiver of the radiation pressure, 
the dust has another role, i.e.,  
the attenuation of ionizing photons. 
To examine its importance, we compare the cases with only the photoionization 
but with (-2Sol\_PI, black dashed) and without dust attenuation (-2Sol\_PI2, dark-blue dashed). 
We can see that, without dust attenuation, the final stellar mass is further reduced from $230~M_{\odot}$ to $180~M_{\odot}$ for this case.


Next, we briefly discuss the cases with lower metallicity, $Z=0$ and $10^{-3}Z_{\odot}$.
Their overall evolution is basically the same as in the case of 
$Z=10^{-2}Z_{\odot}$ described above although the secondary bubble expansion driven by the radiation pressure is not seen in the $Z=0$ case, where 
the dust is absent. In all the cases, the photoionization is the dominant mechanism in regulating the mass accretion onto the star. Contrary to naive expectation from the inflow rate at the end of the collapse phase (Fig. 2 a), the accretion rate onto the protostar is 
actually lower for $Z=0$ and $10^{-3}Z_{\odot}$ than in the $Z=10^{-2} Z_{\odot}$ case early on ($t \lesssim 3 \times 10^4$ years). 
This is because the inner part of the accretion envelope ($M(r) \lesssim 100~\msun$) has higher rotational velocity in those cases (Fig.~\ref{zu_init1}) and the accreted matter accumulates in the disk.
After around the viscous timescale
\begin{align}
t_{\rm vis} &= \frac{1}{3 \pi \alpha} \left( \frac{R}{H}\right)^2 \frac{2 \pi}{\Omega} \nonumber \\
&= 7.5 \times 10^{4} ~ {\rm yr} ~ \left( \frac{\alpha}{1} \right)^{-1} \left( \frac{H/R}{0.1} \right)^{-2} \left( \frac{M_*}{200~M_{\odot}} \right)^{-1/2} \left( \frac{R}{10^4 ~ {\rm au}} \right)^{3/2},
\label{eq:tvis}
\end{align}
the accretion rate rapidly rises and exceeds that for $Z=10^{-2}Z_{\odot}$ (Fig. \ref{zu_stellarevo}) 
as the stationary accretion through the disk establishes. 
The protostar reaches the main sequence by the epoch of $t \sim 0.1$~Myr. 
After that, the star emits ionizing photons at a rate $S_{\rm ion} \sim$ a few $\times 10^{50}~{\rm s^{-1}}$, which is almost the same between the cases with $Z=0$ and $10^{-3}Z_{\odot}$\footnote{While the stellar radius for -3Sol is larger than Z0, the effective temperature is slightly lower. The two effects work oppositely resulting in a similar emissivity of ionizing photons.}. 
The subsequent evolution is also similar between those cases. The stellar mass reaches $\simeq 300~\msun$ at $t \simeq 0.2$~Myr, and the mass growth thereafter is significantly suppressed by the photoionization effect. The final mass is $M_* \simeq 560~\msun$ for 
$Z=0$ and $M_* \simeq 440~\msun$ for $Z=10^{-3}Z_{\odot}$, respectively. 


According to previous studies, the final stellar mass limited by the photoionization is generally higher for higher imposed (i.e., without feedback) accretion rate \citep[e.g.,][]{2014ApJ...781...60H}.
This is consistent with our result in the sense that  
the feedback-limited final mass is the lowest
for $Z = 10^{-2}~Z_\odot$, which has the lowest mean accretion rate before the feedback is taken into account (Fig.~\ref{zu2.1}).
There is, however, some modifications: between cases with 
$Z=0$ and $Z=10^{-3}Z_{\odot}$, the final mass for the former is slightly larger than that for the latter despite with lower mean accretion rate onto the protostar. This is probably caused by the differences in infall rates and rotational velocities in the accretion envelope (see Section \ref{initial_conditions} and Fig.~\ref{zu_init1}). 
For $Z=0$, higher supplies of the mass and angular momentum results in a more massive circumstellar disk.
Since the stellar mass growth in the later phase is largely determined by the amount of gas retained in the disk, the accretion onto the star tends to be prolonged through a more massive disk in cases with the photoionization feedback.
In other words, the final stellar mass is mainly determined by the accretion rate to the disk in very metal-poor environments ($\lesssim 10^{-2}~Z_{\odot}$), where the photoionization is the dominant feedback effect.

\subsection{Accretion evolution from the same envelope structure}
\label{metal_feedback}

 \begin{figure}
 \begin{center}
 \includegraphics[width=\columnwidth]{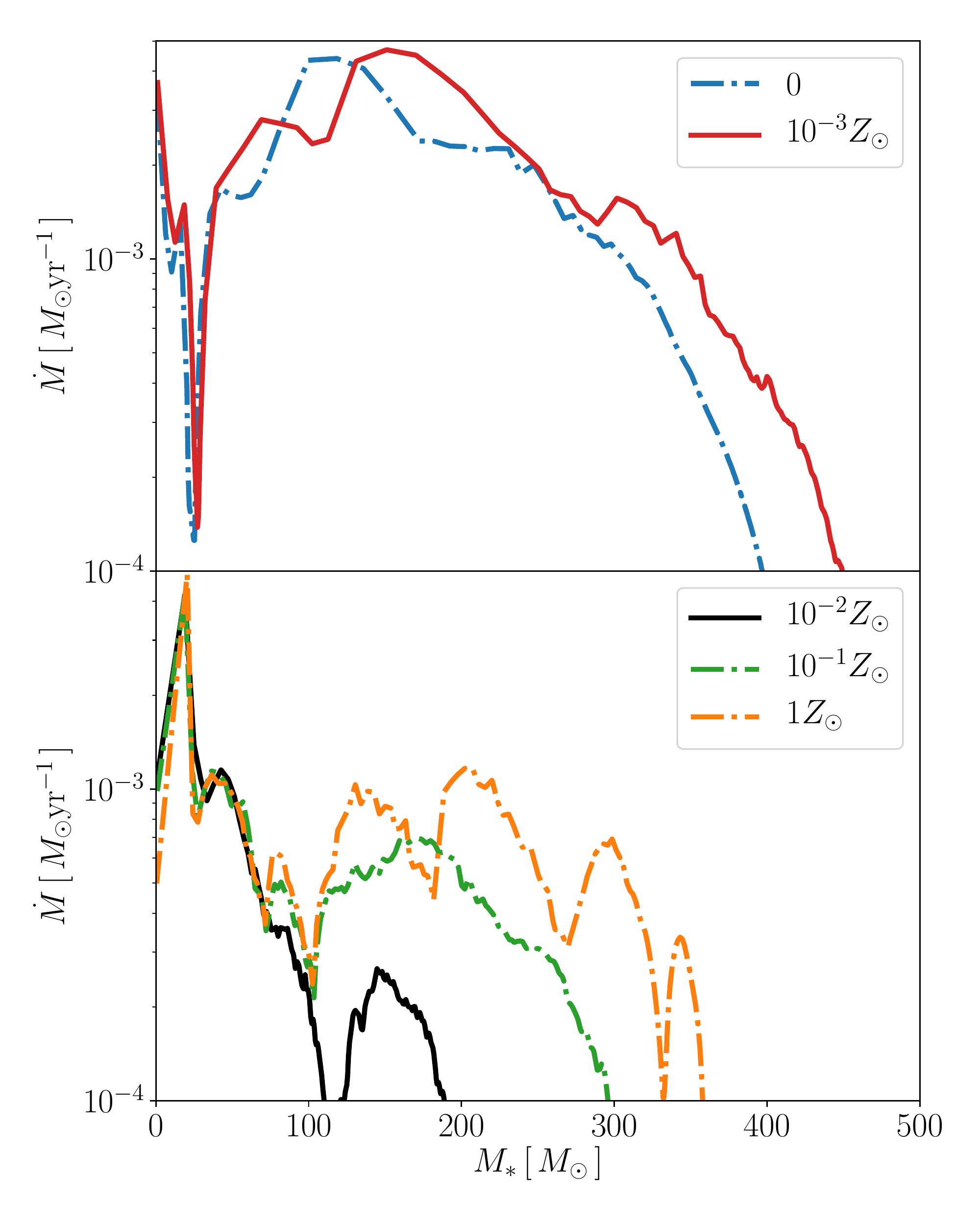}
 \end{center}
 \caption{Metallicity dependence of radiative feedback for a given accretion-envelope structure. {\it Top panel:} the mass accretion histories for metallicities $Z=0$ (Prm\_from-3Sol) and $Z = 10^{-3}~Z_{\odot}$ (-3Sol) from the envelope structure resulting from the cloud collapse with $Z = 10^{-3}~Z_{\odot}$. {\it Bottom panel:} the same as the top panel but from the envelope structure resulting from the collapse with $Z=10^{-2}Z_{\odot}$ for metallicities $Z = 10^{-2}~Z_{\odot}$ (-2Sol), $0.1~Z_\odot$ (-1Sol\_from-2Sol) and $Z = 1~Z_\odot$ (Sol\_from-2Sol).
 }
 \label{zu3.1}
 \end{figure}
 \begin{figure*}
 \begin{center}
 \includegraphics[width=140mm]{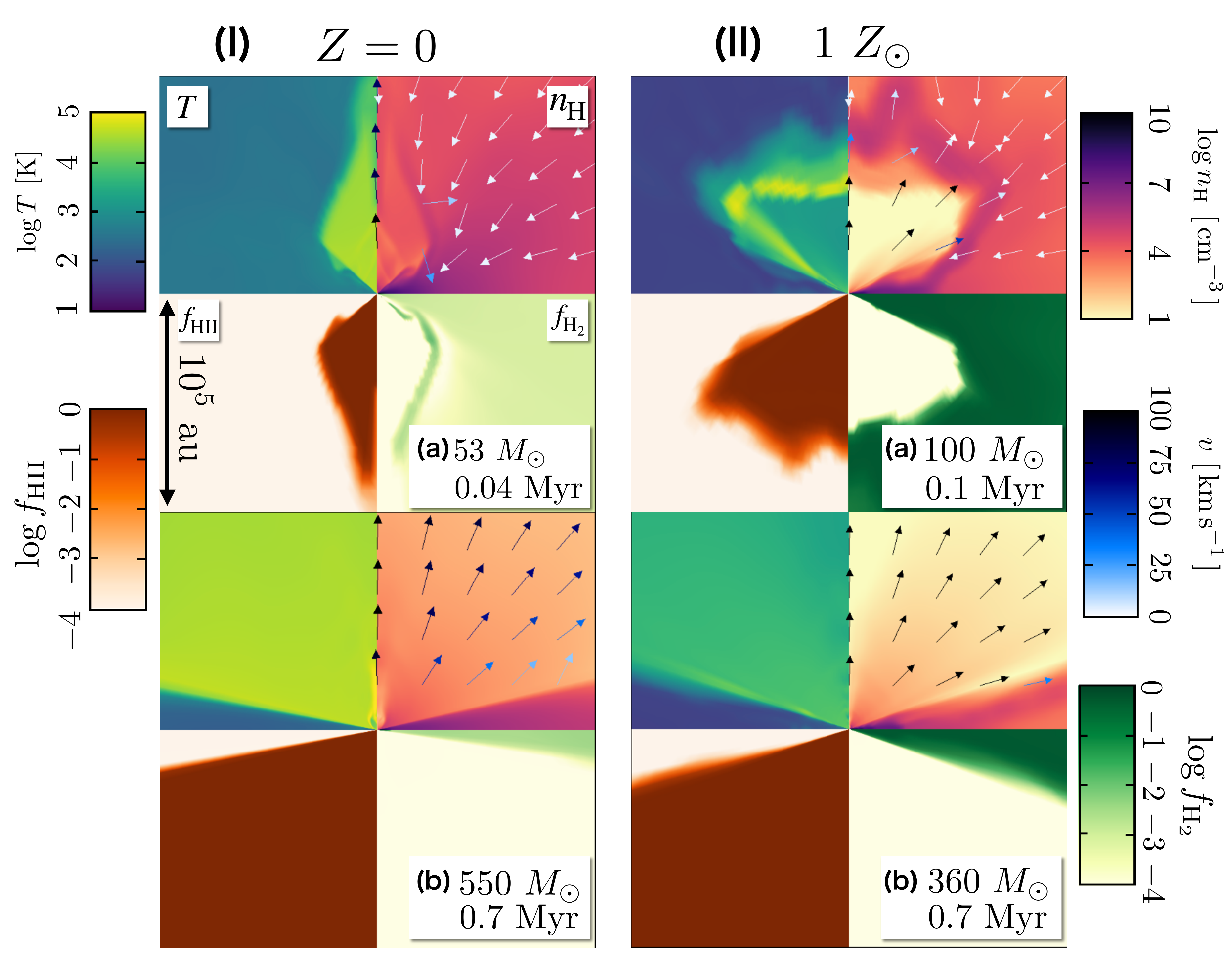}
 \end{center}
 \caption{
Comparison of the accretion envelope structure under radiative feedback between models Prm (left column) and Sol\_from-2Sol (right column), i.e., two extreme and contrasting cases with $Z = 0$ and $1~Z_\odot$.
For each column, the top and bottom panels show snapshots at the different epochs of (a) the polar breakout of the bubble, and (b) almost the end of the stellar mass growth via accretion. 
Each panel illustrates the spatial distribution of the number density $n_{\rm H}$ (upper right), gas temperature $T$ (upper left), abundance of $\rm H_2$ (lower right) and H{\sc ii} (lower left).
}
 \label{zu4}
 \end{figure*}

Since the evolution in the accretion stage is largely affected by the evolution in the collapse phase in the above cases, we focus on the metallicity dependence of the feedback strength in this section.
To extract only metallicity effects on the protostellar accretion, we investigate radiative feedback in cases with different metallicities but from the same envelope structure (see also Sec.~\ref{cases_considerd}).
For this purpose, we assign a certain value of metallicity to the last snapshot at the end of collapse with metallicity $Z=10^{-3}~Z_\odot$ or $10^{-2}~Z_\odot$, maintaining the density and temperature distributions. 
We consider two initial conditions for accretion evolution resulting from the cloud collapse with metallicities $Z=10^{-3}Z_{\odot}$ and $10^{-2}Z_{\odot}$.  
For the initial envelope structure created with $Z=10^{-3}Z_{\odot}$, we study the accretion evolution with $10^{-3}~Z_\odot$ (-3Sol) and $0$ (Prm\_from-3Sol), which is shown in the top panel of Figure \ref{zu3.1}, while for the initial envelope with $Z=10^{-2}Z_{\odot}$ we consider the cases with $10^{-2}~Z_\odot$, $0.1~Z_{\odot}$ and $1~Z_{\odot}$ (cases -1Sol\_from-2Sol and Sol\_from-2Sol) shown in the bottom panel of Figure \ref{zu3.1}. 
We follow the long-term evolution of the protostellar accretion until terminated by radiative feedback.


The final stellar mass for the $Z=0$ case is slightly lower than that for $10^{-3}Z_{\odot}$  (top panel of Figure \ref{zu3.1}).
This trend is opposite to that in Figure~\ref{zu2.1}. 
The discrepancy is due to different rotation velocity profiles for Prm and -3Sol (Fig.~\ref{zu_init1}), which is absent for Prm\_from-3Sol and -3Sol (see also Section~\ref{radiative_feedback}). 


The bottom panel of Figure \ref{zu3.1} shows the mass accretion histories for $Z=10^{-2}Z_{\odot}$ (-2Sol), $10^{-1}Z_{\odot}$ (-1Sol\_from-2Sol) and $Z_{\odot}$ (Sol\_from-2Sol), starting from the envelope structure of $Z=10^{-2}Z_{\odot}$. In the early phase of $M_* \lesssim 100~M_{\odot}$, while radiative feedback is still weak, the accretion histories are mostly determined by the initial condition and very similar to each other. 
At later time, radiative feedback starts operating and the accretion history depends on the metallicity in a way that the final mass is lower, or the feedback is stronger, at lower metallicities. 
This metallicity dependence comes from the dust attenuation of ionizing photons, which can be seen in the following analytic argument. The photoionization feedback quenches the accretion by the photoevaporation of an accretion disk, which proceeds most efficiently near the gravitational radius
\begin{align}
	r_{\rm g} &= \frac{GM_{*}}{c_{\rm i}^2} \nonumber \\
	 &\simeq 1.1 \times 10^3 ~{\rm au} \left( \frac{M_*}{500~M_{\odot}} \right) 
\label{190903.2}
\end{align}
\citep[e.g.,][]{1994ApJ...428..654H,2008ApJ...681..771M}, where $c_{\rm i}$ is the sound speed of the ionized gas for which we assume $T=3 \times 10^{4}~{\rm K}$. Equations~\eqref{eq:rd} and \eqref{190903.2} indicate that the gravitational radius is located outside the dust destruction front and thus the ionizing photons 
around the gravitational radius is attenuated by dust grains.
Using the disk surface density at $r \sim r_{\rm g}$ set by the ionization balance, 
\begin{align}
	n_0 &= \left( \frac{3 S_*}{4 \pi r_{\rm g}^3 \alpha_{\rm B}} \right)^{1/2} \nonumber \\
	&\simeq 2 \times 10^{7} ~{\rm cm^{-3}} \left( \frac{S_*}{10^{51} ~{\rm s^{-1}}} \right)^{1/2} \left( \frac{M_*}{500~M_{\odot}} \right)^{-3/2}, \label{190903.3}
\end{align}
where $S_*$ is the stellar emissivity of ionizing photons, 
we find the optical depth from the star to the gravitational radius, 
\begin{align}
	\tau_{\rm d} &= \rho \kappa_{\rm d} r_{\rm g} \nonumber \\
	&= 3.0 \left( \frac{S_*}{10^{51} ~{\rm s^{-1}}} \right)^{1/2}  \left( \frac{M_*}{200~M_{\odot}} \right)^{-1/2} \left( \frac{\kappa_{\rm d}}{350~{\rm cm^2 g^{-1}}} \right) \left( \frac{Z}{10^{-2}~Z_{\odot}} \right), 
\label{190903.4}
\end{align}
larger than unity for $Z \gtrsim 10^{-2}~Z_{\odot}$, 
where $\kappa_{\rm d}$ is the dust opacity for Lyman-limit frequency at the solar metallicity, meaning that the photoionizing radiation is substantially attenuated by the dust, especially at higher metallicity. 
This explains the trend of weaker feedback at higher metallicity seen in Figure~\ref{zu3.1}.
According to recent studies \citep{2017ApJ...835...32T, 2018ApJ...861...68T, 2018A&A...616A.101K},  
the impact of photoionization is significantly diminished 
by the dust attenuation at the solar metallicity.
In fact, in our cases with $Z=0.1Z_{\odot}$ (-1Sol\_from-2Sol) and $1Z_{\odot}$ 
(Sol\_from-2Sol), the photoionization effect being disabled by this process, the accretion continues until the star becomes very massive with $>300 M_{\odot}$ and the radiation-pressure feedback finally shuts off the inflow.
In this process, the radiation force by the infrared dust re-emission also contributes in terminating the accretion (see Equation \eqref{190903.1}). In particular, it can also repel the flow coming from the shade of the disk, for which the radiation force by the stellar direct light is ineffective \citep[e.g.,][]{Yorke1999, Kuiper2010, 2013ApJ...772...61K}. 
Our results suggest that the metallicity dependence of the radiation-pressure effect is weaker than that of the photoionization owing to the dust attenuation in metal-enriched environments. As a result, the final stellar mass increases with increasing metallicity for $Z >10^{-2}Z_{\odot}$.


Figure \ref{zu4} shows the structure of the accretion envelope and disk at two epochs for the Prm (with $Z=0$ from the envelope structure of $10^{-3}Z_{\odot}$) and Sol\_from-2Sol (with $Z=Z_{\odot}$ from the envelope structure of $10^{-2}Z_{\odot}$) cases. 
Overall evolutionary features in these cases are similar to that in the case with $Z=10^{-2}Z_{\odot}$ (-2Sol) shown in Figure~\ref{zu1}, but with the following notable differences. 
In the case of $Z=0$ (Prm: left panels), secondary bubble, which is 
driven by the radiation force on the dust, does not appear (c.f. Figure~\ref{zu1}-c), and the temperature and density within the \hii region are much higher even in the late stage at $t \simeq 0.7$~Myr. 
In the case of $Z=Z_{\odot}$ (Sol\_from-2Sol: right panels), on the other hand, the secondary bubble starts expanding as early as 0.1~Myr,  simultaneously with the formation of the \hii region.
The density and temperature within the bubble are always low and the outflow velocity is very high. 
From those observations, we can say that 
the case with $Z=10^{-2}Z_{\odot}$ shown in Figure~\ref{zu1}  is intermediate between the cases above in terms of the envelope evolution by radiative feedback although the protostellar accretion is mainly regulated by the photoionization effect as in the $Z=0$ case.

\section{Summary \& Discussion} \label{matome}

We have studied the formation of massive metal-poor stars in a cosmological context. To this end, we have performed a set of radiation-hydrodynamics simulations and followed long-term evolution of the protostellar accretion. We consider both  photoionization and radiation force on dust grains as important feedback mechanisms. 
We have studied the case with $Z=10^{-2}~Z_{\odot}$ (case -2Sol) to examine the role of individual feedback processes. 
In order to clarify the dependence on metallicity, we have also studied the cases with varying metallicity but with using the same envelope structure that is constructed by calculating preceding cloud collapse. 
Our findings are summarized as follows:

\begin{description}
\item[(i)] 
With ($Z \leq 10^{-2}~Z_\odot$), massive stars exceeding $100~\msun$ can form by accretion. 
Photoionization is the dominant mechanism regulating the accretion flow onto the protostar and eventually terminates the mass accretion, similarly in the case of primordial ($Z=0$) star formation. 
At $Z=10^{-2} Z_{\odot}$, radiation force exerted on dust grains drives rapid expansion of a bubble within the \hii region, but its overall impact against the accretion flow is limited. 

\item[(ii)] 
The final stellar mass is higher at lower metallicities, $560~M_{\odot}$ for $Z=0$, $440~M_{\odot}$ for $10^{-3}~Z_{\odot}$, and $190~M_{\odot}$ for $10^{-2}~Z_{\odot}$, respectively.
Although the ionizing photon emissivity at a given stellar mass does not sensitively depend on metallicity, the accretion rate onto the star is smaller at higher metallicities 
because of more efficient cooling in the collapse phase. 
This explains the decrease of the final stellar mass as metallicity increases.

\item[(iii)] 
For a given envelope structure, radiative feedback is stronger at $Z = 10^{-2}~Z_{\odot}$ than at $Z \sim Z_\odot$.
In the ordinary metal-poor range of $Z > 10^{-2}~Z_{\odot}$, ionizing photons are absorbed by dust grains, and photoionization of the accretion disk is suppressed.
The radiation force eventually terminates the mass accretion, but only after the protostar grows to $300~M_{\odot}$.
The final stellar mass increases from $190~M_{\odot}$ for $10^{-2}~Z_{\odot}$ to $360~M_{\odot}$ for $Z_{\odot}$ for our initial conditions.

\end{description}

The points (i) and (ii) above suggest that, even in metal-poor environments at $Z \sim 10^{-2}~Z_{\odot}$, massive stars are formed in a similar manner as in primordial star formation, although with somewhat smaller mass ($190~M_{\odot}$) than the $Z=0$ case $(440~M_{\odot})$. We argue that the metallicity-dependence of the final stellar mass is attributed to the difference in thermal evolution during the early collapse stage at $Z < 10^{-2}~Z_{\odot}$ than by the feedback process. 

The last point (iii) indicates that, at  metallicities $Z \gtrsim 10^{-2}~Z_{\odot}$, very massive star formation is relatively {\it easier} because of the dust attenuation of ionizing photons 
that otherwise effectively halt accretion.
Interestingly, this is opposite to the usual belief that massive star formation is {\it harder} 
if dust grains exist because of effective radiation-pressure feedback. 
Simply, photoionization effect becomes stronger at $Z \sim 10^{-2}~Z_{\odot}$, with less dust attenuation
compared to the case with $Z=Z_{\odot}$.
Observationally, massive stars exceeding $200~M_{\odot}$ exist in the Large Magellanic Cloud, whose average metallicity is $\simeq 0.4~Z_{\odot}$ \citep[e.g.,][]{2010MNRAS.408..731C}.

We discuss an important issue  
regarding disk fragmentation, which our 2D simulations cannot 
properly capture.
It has been shown by 3D simulations that protostellar disks are prone to fragmentation in primordial star formation \citep[e.g.,][]{Machida08, 2012MNRAS.424..399G, 2016ApJ...824..119H} as well as in present-day star formation \citep[e.g.,][]{2017MNRAS.464L..90M, 2018MNRAS.473.3615M}. 
Dust cooling can promote fragmentation with high metallicities \citep{2014MNRAS.439.1884T}. 
As discussed in Section \ref{accretion_to_disk}, the disks with $Z=10^{-3}-10^{-2}~Z_{\odot}$ are more unstable gravitationally than that of $Z=0$.
What happens after fragmentation is still poorly constrained, unfortunately.
In particular, it remains uncertain how many fragments survive without merging with the central star \citep[e.g.,][]{Susa19, Chon19}. Very roughly,. a half the fragments appear to survive in 3D simulations of the primordial star formation and some of them may end up with binaries.
If similar events also occur in very metal-poor environments, massive star binaries would be formed. 
Such low-metallicity binaries,  can be possible progenitors of binary black holes that eventually
become sources of gravitational waves \citep[e.g.,][]{2016PhRvL.116f1102A, 2016ApJ...818L..22A}. 
3D radiation hydrodynamic simulations for those topics are awaited in future studies. 


In the present paper, we have studied radiative feedback from accreting protostars only.
Other effects such as magnetically driven outflows may also contribute to limiting the final stellar mass. Recent magneto-hydrodynamic (MHD) simulations suggest that massive outflow is launched as an outcome of rapid disk accretion \citep[e.g.,][]{2017MNRAS.470.1026M, 2018A&A...620A.182K} and the outflow properties are consistent with observations \citep[e.g.][]{2017NatAs...1E.146H}. 
Other MHD simulations show outflow launching in very metal-poor cases \citep[e.g.,][]{2019MNRAS.486.3741H} although they generally follow
only short-term evolution after the birth of the protostar. 
Further radiation-magneto-hydrodynamic simulations are warranted for future studies on this topic.

\section*{Acknowledgements}
The authors wish to express their cordial thanks to Profs. Masayuki Umemura and Takahiro Tanaka for their continual interest and advice. The authors also thank to Ken Ohsuga and Hidenobu Yajima for useful discussions.
The numerical simulations were performed on the Cray XC40 at the Yukawa Institute Computer Facility and the Cray XC50 (Aterui II) at the Center for Computational Astrophysics (CfCA) of National Astronomical Observatory of Japan. This work is financially supported by the Grants-in-Aid for Basic Research by the Ministry of Education, Science and Culture of Japan (TH: 16H05996, 19H01934,  KO:25287040, 17H01102, 17H02869).
GC is supported by Overseas Research Fellowships
of the Japan Society for the Promotion of Science (JSPS)
for Young Scientists.
RK acknowledges financial support via the Emmy Noether Research Group on Accretion Flows and Feedback in Realistic Models of Massive Star Formation funded by the German Research Foundation (DFG) under grant no. KU 2849/3-1 and KU 2849/3-2.

\section*{DATA AVAILABILITY}
The data underlying this article will be shared on reasonable request to the corresponding author.




\bibliographystyle{mnras}
\bibliography{article} 

\begin{thebibliography}{}
\makeatletter
\relax
\def\mn@urlcharsother{\let\do\@makeother \do\$\do\&\do\#\do\^\do\_\do\%\do\~}
\def\mn@doi{\begingroup\mn@urlcharsother \@ifnextchar [ {\mn@doi@}
  {\mn@doi@[]}}
\def\mn@doi@[#1]#2{\def\@tempa{#1}\ifx\@tempa\@empty \href
  {http://dx.doi.org/#2} {doi:#2}\else \href {http://dx.doi.org/#2} {#1}\fi
  \endgroup}
\def\mn@eprint#1#2{\mn@eprint@#1:#2::\@nil}
\def\mn@eprint@arXiv#1{\href {http://arxiv.org/abs/#1} {{\tt arXiv:#1}}}
\def\mn@eprint@dblp#1{\href {http://dblp.uni-trier.de/rec/bibtex/#1.xml}
  {dblp:#1}}
\def\mn@eprint@#1:#2:#3:#4\@nil{\def\@tempa {#1}\def\@tempb {#2}\def\@tempc
  {#3}\ifx \@tempc \@empty \let \@tempc \@tempb \let \@tempb \@tempa \fi \ifx
  \@tempb \@empty \def\@tempb {arXiv}\fi \@ifundefined
  {mn@eprint@\@tempb}{\@tempb:\@tempc}{\expandafter \expandafter \csname
  mn@eprint@\@tempb\endcsname \expandafter{\@tempc}}}

\bibitem[\protect\citeauthoryear{{Abbott} et~al.,}{{Abbott}
  et~al.}{2016a}]{2016PhRvL.116f1102A}
{Abbott} B.~P.,  et~al., 2016a, \mn@doi [\prl]
  {10.1103/PhysRevLett.116.061102}, \href
  {https://ui.adsabs.harvard.edu/#abs/2016PhRvL.116f1102A} {116, 061102}

\bibitem[\protect\citeauthoryear{{Abbott} et~al.,}{{Abbott}
  et~al.}{2016b}]{2016ApJ...818L..22A}
{Abbott} B.~P.,  et~al., 2016b, \mn@doi [\apj] {10.3847/2041-8205/818/2/L22},
  \href {https://ui.adsabs.harvard.edu/#abs/2016ApJ...818L..22A} {818, L22}

\bibitem[\protect\citeauthoryear{{Abel}, {Anninos}, {Zhang}  \&
  {Norman}}{{Abel} et~al.}{1997}]{1997NewA....2..181A}
{Abel} T.,  {Anninos} P.,  {Zhang} Y.,   {Norman} M.~L.,  1997, \mn@doi [\na]
  {10.1016/S1384-1076(97)00010-9}, \href
  {https://ui.adsabs.harvard.edu/#abs/1997NewA....2..181A} {2, 181}

\bibitem[\protect\citeauthoryear{{Abel}, {Bryan}  \& {Norman}}{{Abel}
  et~al.}{2002}]{2002Sci...295...93A}
{Abel} T.,  {Bryan} G.~L.,   {Norman} M.~L.,  2002, \mn@doi [Science]
  {10.1126/science.295.5552.93}, \href
  {https://ui.adsabs.harvard.edu/abs/2002Sci...295...93A} {295, 93}

\bibitem[\protect\citeauthoryear{{Asano}, {Takeuchi}, {Hirashita}  \&
  {Nozawa}}{{Asano} et~al.}{2013}]{2013MNRAS.432..637A}
{Asano} R.~S.,  {Takeuchi} T.~T.,  {Hirashita} H.,   {Nozawa} T.,  2013,
  \mn@doi [\mnras] {10.1093/mnras/stt506}, \href
  {https://ui.adsabs.harvard.edu/abs/2013MNRAS.432..637A} {432, 637}

\bibitem[\protect\citeauthoryear{{Bromm}, {Ferrara}, {Coppi}  \&
  {Larson}}{{Bromm} et~al.}{2001}]{2001MNRAS.328..969B}
{Bromm} V.,  {Ferrara} A.,  {Coppi} P.~S.,   {Larson} R.~B.,  2001, \mn@doi
  [\mnras] {10.1046/j.1365-8711.2001.04915.x}, \href
  {https://ui.adsabs.harvard.edu/abs/2001MNRAS.328..969B} {328, 969}

\bibitem[\protect\citeauthoryear{{Bromm}, {Coppi}  \& {Larson}}{{Bromm}
  et~al.}{2002}]{Bromm02}
{Bromm} V.,  {Coppi} P.~S.,   {Larson} R.~B.,  2002, \mn@doi [\apj]
  {10.1086/323947}, \href {http://adsabs.harvard.edu/abs/2002ApJ...564...23B}
  {564, 23}

\bibitem[\protect\citeauthoryear{{Cen}}{{Cen}}{1992}]{1992ApJS...78..341C}
{Cen} R.,  1992, \mn@doi [ApJS] {10.1086/191630}, \href
  {https://ui.adsabs.harvard.edu/#abs/1992ApJS...78..341C} {78, 341}

\bibitem[\protect\citeauthoryear{{Chiaki}, {Yoshida}  \& {Hirano}}{{Chiaki}
  et~al.}{2016}]{2016MNRAS.463.2781C}
{Chiaki} G.,  {Yoshida} N.,   {Hirano} S.,  2016, \mn@doi [\mnras]
  {10.1093/mnras/stw2120}, \href
  {https://ui.adsabs.harvard.edu/abs/2016MNRAS.463.2781C} {463, 2781}

\bibitem[\protect\citeauthoryear{{Chiaki}, {Susa}  \& {Hirano}}{{Chiaki}
  et~al.}{2018}]{2018MNRAS.475.4378C}
{Chiaki} G.,  {Susa} H.,   {Hirano} S.,  2018, \mn@doi [\mnras]
  {10.1093/mnras/sty040}, \href
  {https://ui.adsabs.harvard.edu/abs/2018MNRAS.475.4378C} {475, 4378}

\bibitem[\protect\citeauthoryear{{Chon} \& {Hosokawa}}{{Chon} \&
  {Hosokawa}}{2019}]{Chon19}
{Chon} S.,  {Hosokawa} T.,  2019, \mn@doi [\mnras] {10.1093/mnras/stz1824},
  \href {https://ui.adsabs.harvard.edu/abs/2019MNRAS.488.2658C} {488, 2658}

\bibitem[\protect\citeauthoryear{{Clark}, {Glover}, {Smith}, {Greif}, {Klessen}
   \& {Bromm}}{{Clark} et~al.}{2011}]{Clark11}
{Clark} P.~C.,  {Glover} S. C.~O.,  {Smith} R.~J.,  {Greif} T.~H.,  {Klessen}
  R.~S.,   {Bromm} V.,  2011, \mn@doi [Science] {10.1126/science.1198027},
  \href {https://ui.adsabs.harvard.edu/abs/2011Sci...331.1040C} {331, 1040}

\bibitem[\protect\citeauthoryear{{Crowther}, {Schnurr}, {Hirschi}, {Yusof},
  {Parker}, {Goodwin}  \& {Kassim}}{{Crowther}
  et~al.}{2010}]{2010MNRAS.408..731C}
{Crowther} P.~A.,  {Schnurr} O.,  {Hirschi} R.,  {Yusof} N.,  {Parker} R.~J.,
  {Goodwin} S.~P.,   {Kassim} H.~A.,  2010, \mn@doi [\mnras]
  {10.1111/j.1365-2966.2010.17167.x}, \href
  {https://ui.adsabs.harvard.edu/abs/2010MNRAS.408..731C} {408, 731}

\bibitem[\protect\citeauthoryear{{Diaz-Miller}, {Franco}  \&
  {Shore}}{{Diaz-Miller} et~al.}{1998}]{Diaz-Miller98}
{Diaz-Miller} R.~I.,  {Franco} J.,   {Shore} S.~N.,  1998, \mn@doi [\apj]
  {10.1086/305793}, \href
  {https://ui.adsabs.harvard.edu/abs/1998ApJ...501..192D} {501, 192}

\bibitem[\protect\citeauthoryear{{Draine}}{{Draine}}{2011}]{2011piim.book.....D}
{Draine} B.~T.,  2011, {Physics of the Interstellar and Intergalactic Medium}.
{Princeton University Press}

\bibitem[\protect\citeauthoryear{{Draine} \& {Bertoldi}}{{Draine} \&
  {Bertoldi}}{1996}]{1996ApJ...468..269D}
{Draine} B.~T.,  {Bertoldi} F.,  1996, \mn@doi [\apj] {10.1086/177689}, \href
  {https://ui.adsabs.harvard.edu/#abs/1996ApJ...468..269D} {468, 269}

\bibitem[\protect\citeauthoryear{{Forrey}}{{Forrey}}{2013}]{2013ApJ...773L..25F}
{Forrey} R.~C.,  2013, \mn@doi [\apj] {10.1088/2041-8205/773/2/L25}, \href
  {https://ui.adsabs.harvard.edu/#abs/2013ApJ...773L..25F} {773, L25}

\bibitem[\protect\citeauthoryear{{Fukushima}, {Omukai}  \&
  {Hosokawa}}{{Fukushima} et~al.}{2018}]{2018MNRAS.473.4754F}
{Fukushima} H.,  {Omukai} K.,   {Hosokawa} T.,  2018, \mn@doi [\mnras]
  {10.1093/mnras/stx2620}, \href
  {https://ui.adsabs.harvard.edu/#abs/2018MNRAS.473.4754F} {473, 4754}

\bibitem[\protect\citeauthoryear{{Fuller}, {Williams}  \& {Sridharan}}{{Fuller}
  et~al.}{2005}]{2005A&A...442..949F}
{Fuller} G.~A.,  {Williams} S.~J.,   {Sridharan} T.~K.,  2005, \mn@doi
  [Astronomy and Astrophysics] {10.1051/0004-6361:20042110}, \href
  {https://ui.adsabs.harvard.edu/abs/2005A&A...442..949F} {442, 949}

\bibitem[\protect\citeauthoryear{{Galli} \& {Palla}}{{Galli} \&
  {Palla}}{1998}]{1998A&A...335..403G}
{Galli} D.,  {Palla} F.,  1998, \aap, \href
  {https://ui.adsabs.harvard.edu/#abs/1998A&A...335..403G} {335, 403}

\bibitem[\protect\citeauthoryear{{Greif}, {Bromm}, {Clark}, {Glover}, {Smith},
  {Klessen}, {Yoshida}  \& {Springel}}{{Greif}
  et~al.}{2012}]{2012MNRAS.424..399G}
{Greif} T.~H.,  {Bromm} V.,  {Clark} P.~C.,  {Glover} S. C.~O.,  {Smith} R.~J.,
   {Klessen} R.~S.,  {Yoshida} N.,   {Springel} V.,  2012, \mn@doi [\mnras]
  {10.1111/j.1365-2966.2012.21212.x}, \href
  {https://ui.adsabs.harvard.edu/abs/2012MNRAS.424..399G} {424, 399}

\bibitem[\protect\citeauthoryear{{Harries}, {Douglas}  \& {Ali}}{{Harries}
  et~al.}{2017}]{Harries2017}
{Harries} T.~J.,  {Douglas} T.~A.,   {Ali} A.,  2017, \mn@doi [\mnras]
  {10.1093/mnras/stx1490}, \href
  {https://ui.adsabs.harvard.edu/#abs/2017MNRAS.471.4111H} {471, 4111}

\bibitem[\protect\citeauthoryear{{Higuchi}, {Machida}  \& {Susa}}{{Higuchi}
  et~al.}{2019}]{2019MNRAS.486.3741H}
{Higuchi} K.,  {Machida} M.~N.,   {Susa} H.,  2019, \mn@doi [\mnras]
  {10.1093/mnras/stz1079}, \href
  {https://ui.adsabs.harvard.edu/abs/2019MNRAS.486.3741H} {486, 3741}

\bibitem[\protect\citeauthoryear{{Hirano}, {Hosokawa}, {Yoshida}, {Umeda},
  {Omukai}, {Chiaki}  \& {Yorke}}{{Hirano} et~al.}{2014}]{2014ApJ...781...60H}
{Hirano} S.,  {Hosokawa} T.,  {Yoshida} N.,  {Umeda} H.,  {Omukai} K.,
  {Chiaki} G.,   {Yorke} H.~W.,  2014, \mn@doi [\apj]
  {10.1088/0004-637X/781/2/60}, \href
  {https://ui.adsabs.harvard.edu/#abs/2014ApJ...781...60H} {781, 60}

\bibitem[\protect\citeauthoryear{{Hirano}, {Hosokawa}, {Yoshida}, {Omukai}  \&
  {Yorke}}{{Hirano} et~al.}{2015}]{2015MNRAS.448..568H}
{Hirano} S.,  {Hosokawa} T.,  {Yoshida} N.,  {Omukai} K.,   {Yorke} H.~W.,
  2015, \mn@doi [\mnras] {10.1093/mnras/stv044}, \href
  {https://ui.adsabs.harvard.edu/abs/2015MNRAS.448..568H} {448, 568}

\bibitem[\protect\citeauthoryear{{Hirota}, {Machida}, {Matsushita}, {Motogi},
  {Matsumoto}, {Kim}, {Burns}  \& {Honma}}{{Hirota}
  et~al.}{2017}]{2017NatAs...1E.146H}
{Hirota} T.,  {Machida} M.~N.,  {Matsushita} Y.,  {Motogi} K.,  {Matsumoto} N.,
   {Kim} M.~K.,  {Burns} R.~A.,   {Honma} M.,  2017, \mn@doi [Nature Astronomy]
  {10.1038/s41550-017-0146}, \href
  {https://ui.adsabs.harvard.edu/abs/2017NatAs...1E.146H} {1, 0146}

\bibitem[\protect\citeauthoryear{{Hollenbach} \& {McKee}}{{Hollenbach} \&
  {McKee}}{1979}]{1979ApJS...41..555H}
{Hollenbach} D.,  {McKee} C.~F.,  1979, \mn@doi [ApJS] {10.1086/190631}, \href
  {https://ui.adsabs.harvard.edu/#abs/1979ApJS...41..555H} {41, 555}

\bibitem[\protect\citeauthoryear{{Hollenbach} \& {McKee}}{{Hollenbach} \&
  {McKee}}{1989}]{1989ApJ...342..306H}
{Hollenbach} D.,  {McKee} C.~F.,  1989, \mn@doi [\apj] {10.1086/167595}, \href
  {https://ui.adsabs.harvard.edu/#abs/1989ApJ...342..306H} {342, 306}

\bibitem[\protect\citeauthoryear{{Hollenbach}, {Johnstone}, {Lizano}  \&
  {Shu}}{{Hollenbach} et~al.}{1994}]{1994ApJ...428..654H}
{Hollenbach} D.,  {Johnstone} D.,  {Lizano} S.,   {Shu} F.,  1994, \mn@doi
  [\apj] {10.1086/174276}, \href
  {https://ui.adsabs.harvard.edu/#abs/1994ApJ...428..654H} {428, 654}

\bibitem[\protect\citeauthoryear{{Hosokawa} \& {Omukai}}{{Hosokawa} \&
  {Omukai}}{2009a}]{2009ApJ...691..823H}
{Hosokawa} T.,  {Omukai} K.,  2009a, \mn@doi [\apj]
  {10.1088/0004-637X/691/1/823}, \href
  {https://ui.adsabs.harvard.edu/#abs/2009ApJ...691..823H} {691, 823}

\bibitem[\protect\citeauthoryear{{Hosokawa} \& {Omukai}}{{Hosokawa} \&
  {Omukai}}{2009b}]{2009ApJ...703.1810H}
{Hosokawa} T.,  {Omukai} K.,  2009b, \mn@doi [\apj]
  {10.1088/0004-637X/703/2/1810}, \href
  {https://ui.adsabs.harvard.edu/#abs/2009ApJ...703.1810H} {703, 1810}

\bibitem[\protect\citeauthoryear{{Hosokawa}, {Omukai}, {Yoshida}  \&
  {Yorke}}{{Hosokawa} et~al.}{2011}]{2011Sci...334.1250H}
{Hosokawa} T.,  {Omukai} K.,  {Yoshida} N.,   {Yorke} H.~W.,  2011, \mn@doi
  [Science] {10.1126/science.1207433}, \href
  {https://ui.adsabs.harvard.edu/#abs/2011Sci...334.1250H} {334, 1250}

\bibitem[\protect\citeauthoryear{{Hosokawa}, {Hirano}, {Kuiper}, {Yorke},
  {Omukai}  \& {Yoshida}}{{Hosokawa} et~al.}{2016}]{2016ApJ...824..119H}
{Hosokawa} T.,  {Hirano} S.,  {Kuiper} R.,  {Yorke} H.~W.,  {Omukai} K.,
  {Yoshida} N.,  2016, \mn@doi [\apj] {10.3847/0004-637X/824/2/119}, \href
  {https://ui.adsabs.harvard.edu/#abs/2016ApJ...824..119H} {824, 119}

\bibitem[\protect\citeauthoryear{{Isella} \& {Natta}}{{Isella} \&
  {Natta}}{2005}]{2005A&A...438..899I}
{Isella} A.,  {Natta} A.,  2005, \mn@doi [\aap] {10.1051/0004-6361:20052773},
  \href {https://ui.adsabs.harvard.edu/#abs/2005A&A...438..899I} {438, 899}

\bibitem[\protect\citeauthoryear{{Kahn}}{{Kahn}}{1974}]{1974A&A....37..149K}
{Kahn} F.~D.,  1974, \aap, \href
  {https://ui.adsabs.harvard.edu/#abs/1974A&A....37..149K} {37, 149}

\bibitem[\protect\citeauthoryear{{Keto}}{{Keto}}{2003}]{Keto03}
{Keto} E.,  2003, \mn@doi [\apj] {10.1086/379545}, \href
  {https://ui.adsabs.harvard.edu/abs/2003ApJ...599.1196K} {599, 1196}

\bibitem[\protect\citeauthoryear{{Klassen}, {Pudritz}, {Kuiper}, {Peters}  \&
  {Banerjee}}{{Klassen} et~al.}{2016}]{Klassen2016}
{Klassen} M.,  {Pudritz} R.~E.,  {Kuiper} R.,  {Peters} T.,   {Banerjee} R.,
  2016, \mn@doi [\apj] {10.3847/0004-637X/823/1/28}, \href
  {https://ui.adsabs.harvard.edu/#abs/2016ApJ...823...28K} {823, 28}

\bibitem[\protect\citeauthoryear{{K{\"o}lligan} \& {Kuiper}}{{K{\"o}lligan} \&
  {Kuiper}}{2018}]{2018A&A...620A.182K}
{K{\"o}lligan} A.,  {Kuiper} R.,  2018, \mn@doi [\aap]
  {10.1051/0004-6361/201833686}, \href
  {https://ui.adsabs.harvard.edu/abs/2018A&A...620A.182K} {620, A182}

\bibitem[\protect\citeauthoryear{{Kreckel}, {Bruhns}, {{\v{C}}{\'\i}{\v{z}}ek},
  {Glover}, {Miller}, {Urbain}  \& {Savin}}{{Kreckel}
  et~al.}{2010}]{2010Sci...329...69K}
{Kreckel} H.,  {Bruhns} H.,  {{\v{C}}{\'\i}{\v{z}}ek} M.,  {Glover} S.~C.~O.,
  {Miller} K.~A.,  {Urbain} X.,   {Savin} D.~W.,  2010, \mn@doi [Science]
  {10.1126/science.1187191}, \href
  {https://ui.adsabs.harvard.edu/#abs/2010Sci...329...69K} {329, 69}

\bibitem[\protect\citeauthoryear{{Krumholz}}{{Krumholz}}{2018}]{Krumholz18}
{Krumholz} M.~R.,  2018, \mn@doi [Monthly Notices of the Royal Astronomical
  Society] {10.1093/mnras/sty2105}, \href
  {https://ui.adsabs.harvard.edu/abs/2018MNRAS.480.3468K} {480, 3468}

\bibitem[\protect\citeauthoryear{{Krumholz}, {Klein}, {McKee}, {Offner}  \&
  {Cunningham}}{{Krumholz} et~al.}{2009}]{Krumholz2009}
{Krumholz} M.~R.,  {Klein} R.~I.,  {McKee} C.~F.,  {Offner} S. S.~R.,
  {Cunningham} A.~J.,  2009, \mn@doi [Science] {10.1126/science.1165857}, \href
  {https://ui.adsabs.harvard.edu/#abs/2009Sci...323..754K} {323, 754}

\bibitem[\protect\citeauthoryear{{Kuiper} \& {Hosokawa}}{{Kuiper} \&
  {Hosokawa}}{2018}]{2018A&A...616A.101K}
{Kuiper} R.,  {Hosokawa} T.,  2018, \mn@doi [\aap]
  {10.1051/0004-6361/201832638}, \href
  {https://ui.adsabs.harvard.edu/abs/2018A&A...616A.101K} {616, A101}

\bibitem[\protect\citeauthoryear{{Kuiper} \& {Yorke}}{{Kuiper} \&
  {Yorke}}{2013}]{2013ApJ...772...61K}
{Kuiper} R.,  {Yorke} H.~W.,  2013, \mn@doi [\apj]
  {10.1088/0004-637X/772/1/61}, \href
  {https://ui.adsabs.harvard.edu/abs/2013ApJ...772...61K} {772, 61}

\bibitem[\protect\citeauthoryear{{Kuiper}, {Klahr}, {Dullemond}, {Kley}  \&
  {Henning}}{{Kuiper} et~al.}{2010a}]{2010A&A...511A..81K}
{Kuiper} R.,  {Klahr} H.,  {Dullemond} C.,  {Kley} W.,   {Henning} T.,  2010a,
  \mn@doi [\aap] {10.1051/0004-6361/200912355}, \href
  {https://ui.adsabs.harvard.edu/#abs/2010A&A...511A..81K} {511, A81}

\bibitem[\protect\citeauthoryear{{Kuiper}, {Klahr}, {Beuther}  \&
  {Henning}}{{Kuiper} et~al.}{2010b}]{Kuiper2010}
{Kuiper} R.,  {Klahr} H.,  {Beuther} H.,   {Henning} T.,  2010b, \mn@doi [\apj]
  {10.1088/0004-637X/722/2/1556}, \href
  {https://ui.adsabs.harvard.edu/#abs/2010ApJ...722.1556K} {722, 1556}

\bibitem[\protect\citeauthoryear{{Kuiper}, {Klahr}, {Beuther}  \&
  {Henning}}{{Kuiper} et~al.}{2011}]{2011ApJ...732...20K}
{Kuiper} R.,  {Klahr} H.,  {Beuther} H.,   {Henning} T.,  2011, \mn@doi [\apj]
  {10.1088/0004-637X/732/1/20}, \href
  {https://ui.adsabs.harvard.edu/abs/2011ApJ...732...20K} {732, 20}

\bibitem[\protect\citeauthoryear{{Kuiper}, {Klahr}, {Beuther}  \&
  {Henning}}{{Kuiper} et~al.}{2012}]{Kuiper2012}
{Kuiper} R.,  {Klahr} H.,  {Beuther} H.,   {Henning} T.,  2012, \mn@doi [\aap]
  {10.1051/0004-6361/201117808}, \href
  {https://ui.adsabs.harvard.edu/#abs/2012A&A...537A.122K} {537, A122}

\bibitem[\protect\citeauthoryear{{Laor} \& {Draine}}{{Laor} \&
  {Draine}}{1993}]{1993ApJ...402..441L}
{Laor} A.,  {Draine} B.~T.,  1993, \mn@doi [\apj] {10.1086/172149}, \href
  {https://ui.adsabs.harvard.edu/abs/1993ApJ...402..441L} {402, 441}

\bibitem[\protect\citeauthoryear{{Larson}}{{Larson}}{1969}]{1969MNRAS.145..271L}
{Larson} R.~B.,  1969, \mn@doi [\mnras] {10.1093/mnras/145.3.271}, \href
  {https://ui.adsabs.harvard.edu/abs/1969MNRAS.145..271L} {145, 271}

\bibitem[\protect\citeauthoryear{{Larson} \& {Starrfield}}{{Larson} \&
  {Starrfield}}{1971}]{1971A&A....13..190L}
{Larson} R.~B.,  {Starrfield} S.,  1971, \aap, \href
  {https://ui.adsabs.harvard.edu/#abs/1971A&A....13..190L} {13, 190}

\bibitem[\protect\citeauthoryear{{Latif}, {Khochfar}  \& {Whalen}}{{Latif}
  et~al.}{2020}]{2020ApJ...892L...4L}
{Latif} M.~A.,  {Khochfar} S.,   {Whalen} D.,  2020, \mn@doi [\apjl]
  {10.3847/2041-8213/ab7c61}, \href
  {https://ui.adsabs.harvard.edu/abs/2020ApJ...892L...4L} {892, L4}

\bibitem[\protect\citeauthoryear{{Levermore} \& {Pomraning}}{{Levermore} \&
  {Pomraning}}{1981}]{Levermore81}
{Levermore} C.~D.,  {Pomraning} G.~C.,  1981, \mn@doi [\apj] {10.1086/159157},
  \href {https://ui.adsabs.harvard.edu/abs/1981ApJ...248..321L} {248, 321}

\bibitem[\protect\citeauthoryear{{Machida}, {Omukai}, {Matsumoto}  \&
  {Inutsuka}}{{Machida} et~al.}{2008}]{Machida08}
{Machida} M.~N.,  {Omukai} K.,  {Matsumoto} T.,   {Inutsuka} S.-i.,  2008,
  \mn@doi [\apj] {10.1086/533434}, \href
  {https://ui.adsabs.harvard.edu/abs/2008ApJ...677..813M} {677, 813}

\bibitem[\protect\citeauthoryear{{Matsushita}, {Machida}, {Sakurai}  \&
  {Hosokawa}}{{Matsushita} et~al.}{2017}]{2017MNRAS.470.1026M}
{Matsushita} Y.,  {Machida} M.~N.,  {Sakurai} Y.,   {Hosokawa} T.,  2017,
  \mn@doi [\mnras] {10.1093/mnras/stx893}, \href
  {https://ui.adsabs.harvard.edu/abs/2017MNRAS.470.1026M} {470, 1026}

\bibitem[\protect\citeauthoryear{{McKee} \& {Tan}}{{McKee} \&
  {Tan}}{2003}]{2003ApJ...585..850M}
{McKee} C.~F.,  {Tan} J.~C.,  2003, \mn@doi [\apj] {10.1086/346149}, \href
  {https://ui.adsabs.harvard.edu/abs/2003ApJ...585..850M} {585, 850}

\bibitem[\protect\citeauthoryear{{McKee} \& {Tan}}{{McKee} \&
  {Tan}}{2008}]{2008ApJ...681..771M}
{McKee} C.~F.,  {Tan} J.~C.,  2008, \mn@doi [\apj] {10.1086/587434}, \href
  {https://ui.adsabs.harvard.edu/#abs/2008ApJ...681..771M} {681, 771}

\bibitem[\protect\citeauthoryear{{Meyer}, {Vorobyov}, {Kuiper}  \&
  {Kley}}{{Meyer} et~al.}{2017}]{2017MNRAS.464L..90M}
{Meyer} D.~M.~A.,  {Vorobyov} E.~I.,  {Kuiper} R.,   {Kley} W.,  2017, \mn@doi
  [\mnras] {10.1093/mnrasl/slw187}, \href
  {https://ui.adsabs.harvard.edu/abs/2017MNRAS.464L..90M} {464, L90}

\bibitem[\protect\citeauthoryear{{Meyer}, {Kuiper}, {Kley}, {Johnston}  \&
  {Vorobyov}}{{Meyer} et~al.}{2018}]{2018MNRAS.473.3615M}
{Meyer} D.~M.~A.,  {Kuiper} R.,  {Kley} W.,  {Johnston} K.~G.,   {Vorobyov} E.,
   2018, \mn@doi [\mnras] {10.1093/mnras/stx2551}, \href
  {https://ui.adsabs.harvard.edu/abs/2018MNRAS.473.3615M} {473, 3615}

\bibitem[\protect\citeauthoryear{{Mignon-Risse}, {Gonz{\'a}lez},
  {Commer{\c{c}}on}  \& {Rosdahl}}{{Mignon-Risse}
  et~al.}{2020}]{2020A&A...635A..42M}
{Mignon-Risse} R.,  {Gonz{\'a}lez} M.,  {Commer{\c{c}}on} B.,   {Rosdahl} J.,
  2020, \mn@doi [\aap] {10.1051/0004-6361/201936605}, \href
  {https://ui.adsabs.harvard.edu/abs/2020A&A...635A..42M} {635, A42}

\bibitem[\protect\citeauthoryear{{Mignone}, {Bodo}, {Massaglia}, {Matsakos},
  {Tesileanu}, {Zanni}  \& {Ferrari}}{{Mignone}
  et~al.}{2007}]{2007ApJS..170..228M}
{Mignone} A.,  {Bodo} G.,  {Massaglia} S.,  {Matsakos} T.,  {Tesileanu} O.,
  {Zanni} C.,   {Ferrari} A.,  2007, \mn@doi [ApJS] {10.1086/513316}, \href
  {https://ui.adsabs.harvard.edu/#abs/2007ApJS..170..228M} {170, 228}

\bibitem[\protect\citeauthoryear{{Nakatani} \& {Yoshida}}{{Nakatani} \&
  {Yoshida}}{2019}]{2019ApJ...883..127N}
{Nakatani} R.,  {Yoshida} N.,  2019, \mn@doi [\apj] {10.3847/1538-4357/ab380a},
  \href {https://ui.adsabs.harvard.edu/abs/2019ApJ...883..127N} {883, 127}

\bibitem[\protect\citeauthoryear{{Nakatani}, {Hosokawa}, {Yoshida}, {Nomura}
  \& {Kuiper}}{{Nakatani} et~al.}{2018a}]{2018ApJ...857...57N}
{Nakatani} R.,  {Hosokawa} T.,  {Yoshida} N.,  {Nomura} H.,   {Kuiper} R.,
  2018a, \mn@doi [\apj] {10.3847/1538-4357/aab70b}, \href
  {https://ui.adsabs.harvard.edu/#abs/2018ApJ...857...57N} {857, 57}

\bibitem[\protect\citeauthoryear{{Nakatani}, {Hosokawa}, {Yoshida}, {Nomura}
  \& {Kuiper}}{{Nakatani} et~al.}{2018b}]{2018ApJ...865...75N}
{Nakatani} R.,  {Hosokawa} T.,  {Yoshida} N.,  {Nomura} H.,   {Kuiper} R.,
  2018b, \mn@doi [\apj] {10.3847/1538-4357/aad9fd}, \href
  {https://ui.adsabs.harvard.edu/abs/2018ApJ...865...75N} {865, 75}

\bibitem[\protect\citeauthoryear{{Nozawa} et~al.,}{{Nozawa}
  et~al.}{2008}]{2008ApJ...684.1343N}
{Nozawa} T.,  et~al., 2008, \mn@doi [\apj] {10.1086/589961}, \href
  {https://ui.adsabs.harvard.edu/abs/2008ApJ...684.1343N} {684, 1343}

\bibitem[\protect\citeauthoryear{{Nussbaumer} \& {Storey}}{{Nussbaumer} \&
  {Storey}}{1983}]{1983A&A...126...75N}
{Nussbaumer} H.,  {Storey} P.~J.,  1983, \aap, \href
  {https://ui.adsabs.harvard.edu/#abs/1983A&A...126...75N} {126, 75}

\bibitem[\protect\citeauthoryear{{Omukai}}{{Omukai}}{2000}]{2000ApJ...534..809O}
{Omukai} K.,  2000, \mn@doi [\apj] {10.1086/308776}, \href
  {https://ui.adsabs.harvard.edu/#abs/2000ApJ...534..809O} {534, 809}

\bibitem[\protect\citeauthoryear{{Omukai} \& {Inutsuka}}{{Omukai} \&
  {Inutsuka}}{2002}]{2002MNRAS.332...59O}
{Omukai} K.,  {Inutsuka} S.-i.,  2002, \mn@doi [\mnras]
  {10.1046/j.1365-8711.2002.05276.x}, \href
  {https://ui.adsabs.harvard.edu/#abs/2002MNRAS.332...59O} {332, 59}

\bibitem[\protect\citeauthoryear{{Omukai} \& {Nishi}}{{Omukai} \&
  {Nishi}}{1998}]{1998ApJ...508..141O}
{Omukai} K.,  {Nishi} R.,  1998, \mn@doi [\apj] {10.1086/306395}, \href
  {https://ui.adsabs.harvard.edu/abs/1998ApJ...508..141O} {508, 141}

\bibitem[\protect\citeauthoryear{{Omukai} \& {Palla}}{{Omukai} \&
  {Palla}}{2003}]{2003ApJ...589..677O}
{Omukai} K.,  {Palla} F.,  2003, \mn@doi [\apj] {10.1086/374810}, \href
  {https://ui.adsabs.harvard.edu/#abs/2003ApJ...589..677O} {589, 677}

\bibitem[\protect\citeauthoryear{{Omukai}, {Tsuribe}, {Schneider}  \&
  {Ferrara}}{{Omukai} et~al.}{2005}]{2005ApJ...626..627O}
{Omukai} K.,  {Tsuribe} T.,  {Schneider} R.,   {Ferrara} A.,  2005, \mn@doi
  [\apj] {10.1086/429955}, \href
  {https://ui.adsabs.harvard.edu/#abs/2005ApJ...626..627O} {626, 627}

\bibitem[\protect\citeauthoryear{{Osterbrock}}{{Osterbrock}}{1989}]{1989agna.book.....O}
{Osterbrock} D.~E.,  1989, {Astrophysics of gaseous nebulae and active galactic
  nuclei}.
{University Science Books}

\bibitem[\protect\citeauthoryear{{Osterbrock} \& {Ferland}}{{Osterbrock} \&
  {Ferland}}{2006}]{2006agna.book.....O}
{Osterbrock} D.~E.,  {Ferland} G.~J.,  2006, {Astrophysics of gaseous nebulae
  and active galactic nuclei}.
{University Science Books}

\bibitem[\protect\citeauthoryear{{Palla}, {Salpeter}  \& {Stahler}}{{Palla}
  et~al.}{1983}]{1983ApJ...271..632P}
{Palla} F.,  {Salpeter} E.~E.,   {Stahler} S.~W.,  1983, \mn@doi [\apj]
  {10.1086/161231}, \href
  {https://ui.adsabs.harvard.edu/#abs/1983ApJ...271..632P} {271, 632}

\bibitem[\protect\citeauthoryear{{Pequignot}, {Petitjean}  \&
  {Boisson}}{{Pequignot} et~al.}{1991}]{1991A&A...251..680P}
{Pequignot} D.,  {Petitjean} P.,   {Boisson} C.,  1991, \aap, \href
  {https://ui.adsabs.harvard.edu/#abs/1991A&A...251..680P} {251, 680}

\bibitem[\protect\citeauthoryear{{Peters}, {Klessen}, {Mac Low}  \&
  {Banerjee}}{{Peters} et~al.}{2010}]{2010ApJ...725..134P}
{Peters} T.,  {Klessen} R.~S.,  {Mac Low} M.-M.,   {Banerjee} R.,  2010,
  \mn@doi [The Astrophysical Journal] {10.1088/0004-637X/725/1/134}, \href
  {https://ui.adsabs.harvard.edu/abs/2010ApJ...725..134P} {725, 134}

\bibitem[\protect\citeauthoryear{{Peters}, {Banerjee}, {Klessen}  \& {Mac
  Low}}{{Peters} et~al.}{2011}]{2011ApJ...729...72P}
{Peters} T.,  {Banerjee} R.,  {Klessen} R.~S.,   {Mac Low} M.-M.,  2011,
  \mn@doi [The Astrophysical Journal] {10.1088/0004-637X/729/1/72}, \href
  {https://ui.adsabs.harvard.edu/abs/2011ApJ...729...72P} {729, 72}

\bibitem[\protect\citeauthoryear{{Pollack}, {Hollenbach}, {Beckwith},
  {Simonelli}, {Roush}  \& {Fong}}{{Pollack}
  et~al.}{1994}]{1994ApJ...421..615P}
{Pollack} J.~B.,  {Hollenbach} D.,  {Beckwith} S.,  {Simonelli} D.~P.,  {Roush}
  T.,   {Fong} W.,  1994, \mn@doi [\apj] {10.1086/173677}, \href
  {https://ui.adsabs.harvard.edu/#abs/1994ApJ...421..615P} {421, 615}

\bibitem[\protect\citeauthoryear{{R{\'e}my-Ruyer} et~al.,}{{R{\'e}my-Ruyer}
  et~al.}{2014}]{2014A&A...563A..31R}
{R{\'e}my-Ruyer} A.,  et~al., 2014, \mn@doi [\aap]
  {10.1051/0004-6361/201322803}, \href
  {https://ui.adsabs.harvard.edu/abs/2014A&A...563A..31R} {563, A31}

\bibitem[\protect\citeauthoryear{{Rosen}, {Krumholz}, {McKee}  \&
  {Klein}}{{Rosen} et~al.}{2016}]{Rosen2016}
{Rosen} A.~L.,  {Krumholz} M.~R.,  {McKee} C.~F.,   {Klein} R.~I.,  2016,
  \mn@doi [\mnras] {10.1093/mnras/stw2153}, \href
  {https://ui.adsabs.harvard.edu/#abs/2016MNRAS.463.2553R} {463, 2553}

\bibitem[\protect\citeauthoryear{{Santoro} \& {Shull}}{{Santoro} \&
  {Shull}}{2006}]{2006ApJ...643...26S}
{Santoro} F.,  {Shull} J.~M.,  2006, \mn@doi [\apj] {10.1086/501518}, \href
  {https://ui.adsabs.harvard.edu/#abs/2006ApJ...643...26S} {643, 26}

\bibitem[\protect\citeauthoryear{{Sartorio}, {Vandenbroucke},
  {Falceta-Goncalves}, {Wood}  \& {Keto}}{{Sartorio}
  et~al.}{2019}]{2019MNRAS.486.5171S}
{Sartorio} N.~S.,  {Vandenbroucke} B.,  {Falceta-Goncalves} D.,  {Wood} K.,
  {Keto} E.,  2019, \mn@doi [\mnras] {10.1093/mnras/stz1187}, \href
  {https://ui.adsabs.harvard.edu/abs/2019MNRAS.486.5171S} {486, 5171}

\bibitem[\protect\citeauthoryear{{Schneider}, {Ferrara}, {Salvaterra}, {Omukai}
   \& {Bromm}}{{Schneider} et~al.}{2003}]{2003Natur.422..869S}
{Schneider} R.,  {Ferrara} A.,  {Salvaterra} R.,  {Omukai} K.,   {Bromm} V.,
  2003, \mn@doi [\nat] {10.1038/nature01579}, \href
  {https://ui.adsabs.harvard.edu/abs/2003Natur.422..869S} {422, 869}

\bibitem[\protect\citeauthoryear{{Shakura} \& {Sunyaev}}{{Shakura} \&
  {Sunyaev}}{1973}]{1973A&A....24..337S}
{Shakura} N.~I.,  {Sunyaev} R.~A.,  1973, \aap, \href
  {https://ui.adsabs.harvard.edu/abs/1973A&A....24..337S} {500, 33}

\bibitem[\protect\citeauthoryear{{Shapiro} \& {Kang}}{{Shapiro} \&
  {Kang}}{1987}]{1987ApJ...318...32S}
{Shapiro} P.~R.,  {Kang} H.,  1987, \mn@doi [\apj] {10.1086/165350}, \href
  {https://ui.adsabs.harvard.edu/#abs/1987ApJ...318...32S} {318, 32}

\bibitem[\protect\citeauthoryear{{Spitzer}}{{Spitzer}}{1978}]{1978ppim.book.....S}
{Spitzer} L.,  1978, {Physical processes in the interstellar medium},
  \mn@doi{10.1002/9783527617722.
}

\bibitem[\protect\citeauthoryear{{Springel}}{{Springel}}{2005}]{2005MNRAS.364.1105S}
{Springel} V.,  2005, \mn@doi [\mnras] {10.1111/j.1365-2966.2005.09655.x},
  \href {https://ui.adsabs.harvard.edu/abs/2005MNRAS.364.1105S} {364, 1105}

\bibitem[\protect\citeauthoryear{{Stacy}, {Greif}  \& {Bromm}}{{Stacy}
  et~al.}{2010}]{Stacy10}
{Stacy} A.,  {Greif} T.~H.,   {Bromm} V.,  2010, \mn@doi [\mnras]
  {10.1111/j.1365-2966.2009.16113.x}, \href
  {https://ui.adsabs.harvard.edu/abs/2010MNRAS.403...45S} {403, 45}

\bibitem[\protect\citeauthoryear{{Stacy}, {Bromm}  \& {Lee}}{{Stacy}
  et~al.}{2016}]{2016MNRAS.462.1307S}
{Stacy} A.,  {Bromm} V.,   {Lee} A.~T.,  2016, \mn@doi [\mnras]
  {10.1093/mnras/stw1728}, \href
  {https://ui.adsabs.harvard.edu/abs/2016MNRAS.462.1307S} {462, 1307}

\bibitem[\protect\citeauthoryear{{Susa}}{{Susa}}{2019}]{Susa19}
{Susa} H.,  2019, \mn@doi [\apj] {10.3847/1538-4357/ab1b6f}, \href
  {https://ui.adsabs.harvard.edu/abs/2019ApJ...877...99S} {877, 99}

\bibitem[\protect\citeauthoryear{{Susa}, {Hasegawa}  \& {Tominaga}}{{Susa}
  et~al.}{2014}]{2014ApJ...792...32S}
{Susa} H.,  {Hasegawa} K.,   {Tominaga} N.,  2014, \mn@doi [\apj]
  {10.1088/0004-637X/792/1/32}, \href
  {https://ui.adsabs.harvard.edu/abs/2014ApJ...792...32S} {p.~32}

\bibitem[\protect\citeauthoryear{{Takahashi}, {Silk}  \&
  {Hollenbach}}{{Takahashi} et~al.}{1983}]{1983ApJ...275..145T}
{Takahashi} T.,  {Silk} J.,   {Hollenbach} D.~J.,  1983, \mn@doi [\apj]
  {10.1086/161521}, \href
  {https://ui.adsabs.harvard.edu/#abs/1983ApJ...275..145T} {275, 145}

\bibitem[\protect\citeauthoryear{{Takahashi}, {Inutsuka}  \&
  {Machida}}{{Takahashi} et~al.}{2013}]{2013ApJ...770...71T}
{Takahashi} S.~Z.,  {Inutsuka} S.-i.,   {Machida} M.~N.,  2013, \mn@doi [\apj]
  {10.1088/0004-637X/770/1/71}, \href
  {https://ui.adsabs.harvard.edu/#abs/2013ApJ...770...71T} {770, 71}

\bibitem[\protect\citeauthoryear{{Tanaka} \& {Omukai}}{{Tanaka} \&
  {Omukai}}{2014}]{2014MNRAS.439.1884T}
{Tanaka} K. E.~I.,  {Omukai} K.,  2014, \mn@doi [\mnras]
  {10.1093/mnras/stu069}, \href
  {https://ui.adsabs.harvard.edu/abs/2014MNRAS.439.1884T} {439, 1884}

\bibitem[\protect\citeauthoryear{{Tanaka}, {Nakamoto}  \& {Omukai}}{{Tanaka}
  et~al.}{2013}]{2013ApJ...773..155T}
{Tanaka} K. E.~I.,  {Nakamoto} T.,   {Omukai} K.,  2013, \mn@doi [\apj]
  {10.1088/0004-637X/773/2/155}, \href
  {https://ui.adsabs.harvard.edu/abs/2013ApJ...773..155T} {773, 155}

\bibitem[\protect\citeauthoryear{{Tanaka}, {Tan}  \& {Zhang}}{{Tanaka}
  et~al.}{2017}]{2017ApJ...835...32T}
{Tanaka} K. E.~I.,  {Tan} J.~C.,   {Zhang} Y.,  2017, \mn@doi [\apj]
  {10.3847/1538-4357/835/1/32}, \href
  {https://ui.adsabs.harvard.edu/abs/2017ApJ...835...32T} {835, 32}

\bibitem[\protect\citeauthoryear{{Tanaka}, {Tan}, {Zhang}  \&
  {Hosokawa}}{{Tanaka} et~al.}{2018}]{2018ApJ...861...68T}
{Tanaka} K. E.~I.,  {Tan} J.~C.,  {Zhang} Y.,   {Hosokawa} T.,  2018, \mn@doi
  [\apj] {10.3847/1538-4357/aac892}, \href
  {https://ui.adsabs.harvard.edu/#abs/2018ApJ...861...68T} {861, 68}

\bibitem[\protect\citeauthoryear{{Tielens} \& {Hollenbach}}{{Tielens} \&
  {Hollenbach}}{1985}]{1985ApJ...291..722T}
{Tielens} A.~G.~G.~M.,  {Hollenbach} D.,  1985, \mn@doi [\apj]
  {10.1086/163111}, \href
  {https://ui.adsabs.harvard.edu/#abs/1985ApJ...291..722T} {291, 722}

\bibitem[\protect\citeauthoryear{{Toomre}}{{Toomre}}{1964}]{1964ApJ...139.1217T}
{Toomre} A.,  1964, \mn@doi [\apj] {10.1086/147861}, \href
  {https://ui.adsabs.harvard.edu/#abs/1964ApJ...139.1217T} {139, 1217}

\bibitem[\protect\citeauthoryear{{Voronov}}{{Voronov}}{1997}]{1997ADNDT..65....1V}
{Voronov} G.~S.,  1997, \mn@doi [Atomic Data and Nuclear Data Tables]
  {10.1006/adnd.1997.0732}, \href
  {https://ui.adsabs.harvard.edu/#abs/1997ADNDT..65....1V} {65, 1}

\bibitem[\protect\citeauthoryear{{Wolfire} \& {Cassinelli}}{{Wolfire} \&
  {Cassinelli}}{1987}]{1987ApJ...319..850W}
{Wolfire} M.~G.,  {Cassinelli} J.~P.,  1987, \mn@doi [\apj] {10.1086/165503},
  \href {https://ui.adsabs.harvard.edu/#abs/1987ApJ...319..850W} {319, 850}

\bibitem[\protect\citeauthoryear{{Wyrowski} et~al.,}{{Wyrowski}
  et~al.}{2016}]{2016A&A...585A.149W}
{Wyrowski} F.,  et~al., 2016, \mn@doi [Astronomy and Astrophysics]
  {10.1051/0004-6361/201526361}, \href
  {https://ui.adsabs.harvard.edu/abs/2016A&A...585A.149W} {585, A149}

\bibitem[\protect\citeauthoryear{{Yorke} \& {Bodenheimer}}{{Yorke} \&
  {Bodenheimer}}{1999}]{Yorke1999}
{Yorke} H.~W.,  {Bodenheimer} P.,  1999, \mn@doi [\apj] {10.1086/307867}, \href
  {https://ui.adsabs.harvard.edu/#abs/1999ApJ...525..330Y} {525, 330}

\bibitem[\protect\citeauthoryear{{Yorke} \& {Kruegel}}{{Yorke} \&
  {Kruegel}}{1977}]{1977A&A....54..183Y}
{Yorke} H.~W.,  {Kruegel} E.,  1977, \aap, \href
  {https://ui.adsabs.harvard.edu/#abs/1977A&A....54..183Y} {54, 183}

\bibitem[\protect\citeauthoryear{{Yorke} \& {Sonnhalter}}{{Yorke} \&
  {Sonnhalter}}{2002}]{Yorke2002}
{Yorke} H.~W.,  {Sonnhalter} C.,  2002, \mn@doi [\apj] {10.1086/339264}, \href
  {https://ui.adsabs.harvard.edu/#abs/2002ApJ...569..846Y} {569, 846}

\bibitem[\protect\citeauthoryear{{Yoshida}, {Abel}, {Hernquist}  \&
  {Sugiyama}}{{Yoshida} et~al.}{2003}]{Y03}
{Yoshida} N.,  {Abel} T.,  {Hernquist} L.,   {Sugiyama} N.,  2003, \mn@doi
  [\apj] {10.1086/375810}, \href
  {http://adsabs.harvard.edu/abs/2003ApJ...592..645Y} {592, 645}

\bibitem[\protect\citeauthoryear{{Yoshida}, {Omukai}, {Hernquist}  \&
  {Abel}}{{Yoshida} et~al.}{2006}]{2006ApJ...652....6Y}
{Yoshida} N.,  {Omukai} K.,  {Hernquist} L.,   {Abel} T.,  2006, \mn@doi [\apj]
  {10.1086/507978}, \href
  {https://ui.adsabs.harvard.edu/abs/2006ApJ...652....6Y} {652, 6}

\bibitem[\protect\citeauthoryear{{Yoshida}, {Omukai}  \& {Hernquist}}{{Yoshida}
  et~al.}{2008}]{2008Sci...321..669Y}
{Yoshida} N.,  {Omukai} K.,   {Hernquist} L.,  2008, \mn@doi [Science]
  {10.1126/science.1160259}, \href
  {https://ui.adsabs.harvard.edu/abs/2008Sci...321..669Y} {321, 669}

\bibitem[\protect\citeauthoryear{{Zhu}, {Hartmann}  \& {Gammie}}{{Zhu}
  et~al.}{2010}]{2010ApJ...713.1143Z}
{Zhu} Z.,  {Hartmann} L.,   {Gammie} C.,  2010, \mn@doi [\apj]
  {10.1088/0004-637X/713/2/1143}, \href
  {https://ui.adsabs.harvard.edu/#abs/2010ApJ...713.1143Z} {713, 1143}

\bibitem[\protect\citeauthoryear{{Zinnecker} \& {Yorke}}{{Zinnecker} \&
  {Yorke}}{2007}]{ZY07}
{Zinnecker} H.,  {Yorke} H.~W.,  2007, \mn@doi [\araa]
  {10.1146/annurev.astro.44.051905.092549}, \href
  {https://ui.adsabs.harvard.edu/abs/2007ARA&A..45..481Z} {45, 481}

\makeatother
\end{thebibliography}




\appendix
\section{Numerical Method}\label{chemapp}
\subsection{Chemical Networks}

In Table \ref{chem_process}, we summarize the rates of chemical reactions relevant with ${\rm H}, {\rm H_{2}}, {\rm H^{-}}, {\rm H^{+}}$ and ${\rm e}$ incorporated in our RHD simulations.
We also consider $\rm C^{+}$, O, $\rm O^{+}$, and $\rm O^{2+}$ as components of heavy elements.

\begin{table*}
 	\caption{Chemical Reactions }
 	\label{chem_process}
 	\centering
 	\begin{tabular}{lllc}
 		\hline \hline
 		Number & Reaction & Rate Coefficient & Reference \\
 		\hline
 		${\rm H1}$ & ${\rm H^{+}} + e \rightarrow {\rm H} + \gamma$ & $k_{\rm H1} = 2.753  \times 10^{-14} \left( 315614 / T \right)^{1.5} \left[1 + \left( 115188 / T \right)^{0.407} \right]^{-2.242} ({\rm case \, B}) $& 1\\

  		${\rm H2}$ & ${\rm H} + e \rightarrow {\rm H^{-}} + \gamma$ & $k_{\rm H2} = 1.4 \times 10^{-18}T^{0.928} \exp(-T/1.62 \times 10^{4})$ & 2 \\
 		${\rm H3}$ & ${\rm H^{-}} + {\rm H} \rightarrow {\rm H_{2}} + {\rm e} $ & $k_{\rm H3} = 1.35 \times 10^{-9} \left[ T^{0.098493} + 0.32852 T^{0.5561} + 2.771 \times 10^{-7} T^{2.1826} \right]$ & 6 \\
 		& & $/ \left[ 1 + 6.191 \times 10^{-3} T^{1.0461} + 8.9712 \times 10^{-11} T^{3.0424} + 3.2576 \times 10^{-14} T^{3.7741}  \right]$ & \\
 		${\rm H4}$ & ${\rm H_{2}} + {\rm H} \rightarrow 3 {\rm H} $ & $k_{\rm H4} = k_{\rm H}^{1-a} k_{\rm L}^a$ & 3 \\
 		& & \hspace{2mm} $k_{\rm L} = 1.12 \times 10^{-10} \exp(-7.035 \times 10^{4} / T)$ & \\
 		& & \hspace{2mm} $k_{\rm H} = 6.5 \times 10^{-7} T^{-1/2} \exp(-5.2 \times 10^{4} / T) \left[ 1 - \exp(-6000/T) \right]$ & \\
 		& & \hspace{2mm} $a = \left( 1 + n / n_{\rm cr} \right)^{-1}$ & \\
 		& & \hspace{2mm} $\log_{10} \left(n_{\rm cr}\right) = 4.0 - 0.416 \log_{10} \left( T / 10^{4}  \right) - 0.327 \left[ \log_{10} \left( T / 10^{4}  \right) \right]^2$ & \\
 		${\rm H5}$ & $ 3 {\rm H} \rightarrow  {\rm H_{2}} + {\rm H} $ & $k_{\rm H5} =  6 \times 10^{-32} T^{-0.25} + 2 \times 10^{-31} T^{-0.5}$ & 4 \\
 		${\rm H6}$ & $2 {\rm H} + {\rm H_{2}} \rightarrow 2 {\rm H_{2}}$ & $k_{\rm H6} = k_{\rm H5} / 8$ & 5 \\
 		${\rm H7}$ & $2 {\rm H_{2}} \rightarrow 2 {\rm H} + {\rm H_{2}}$ & $k_{\rm H7} = k_{\rm high}^{1 - a} k_{\rm low}^{a}$ & 5 \\
 		& & \hspace{2mm} $k_{\rm low} = 1.18 \times 10^{-10} \exp(-6.95 \times 10^{4} / T)$ &  \\
 		& & \hspace{2mm} $k_{\rm high} = 8.125 \times 10^{-8} T^{-1/2} \exp(-5.2 \times 10^{4} / T) \left[ 1 - \exp(-6000/T) \right]$ & \\
 		& & \hspace{2mm} $a = \left( 1 + n / n_{\rm cr} \right)^{-1}$ & \\
 		& & \hspace{2mm} $\log_{10} (n_{\rm cr}) = 4.845 - 1.3 \log_{10} \left(T / 10^{4} \right) + 1.62 \left[ \log_{10} \left( T / 10^{4}  \right) \right]^2$ & \\
		${\rm H8}$ & ${\rm 2 H } + {\rm grain} \rightarrow {\rm H_{2}} $ &  $k_{\rm H8} = 6.0 \times 10^{-17} \sqrt{T/300} f_{a} \left(Z/Z_{\odot} \right) [1.0 + 4.0 \times 10^{-2} \sqrt{T+T_{\rm gr}} + 2.0 \times 10^{-3} T + 8.0 \times 10^{-6} T^{2}]^{-1}$ &  8\\
		& & \hspace{2mm} $f_{a} = [1.0 + \exp (7.5 \times 10^{2} (1/75 - T_{\rm gr}^{-1}))]^{-1}$ & \\
		${\rm H9}$ & ${\rm H} + {\rm e} \rightarrow {\rm H^{+}} + 2 {\rm e}$ & $k_{\rm H9} = \exp [ - 32.71396786 + 13.536556 \times  \left( \ln T \left({\rm ev} \right) \right) -5.73932875 \left( \ln T \left({\rm ev} \right) \right)^{2} +1.56315498  \left( \ln T \left({\rm ev} \right) \right)^{3} $ & 7 \\
		& & \hspace{0.5cm} $-0.2877056   \left( \ln T \left({\rm ev} \right) \right)^{4} + 3.48255977 \times 10^{-2}  \left( \ln T \left({\rm ev} \right) \right)^{5} -2.63197617 \times 10^{-3}  \left( \ln T \left({\rm ev} \right) \right)^{6}$ & \\
		&& \hspace{0.5cm}$ + 1.11954395 \times 10^{-4}  \left( \ln T \left({\rm ev} \right) \right)^{7} - 2.03914985 \times 10^{-6}  \left( \ln T \left({\rm ev} \right) \right)^{8} $ & \\%
		${\rm H11}$ & ${\rm H_{2}} + {\rm e} \rightarrow 2 {\rm H} + {\rm e}$ & $k_{\rm H11} = 4.4 \times 10^{-10} T^{0.35} \exp(-1.02 \times 10^{5} / T)$ \hspace{5cm} & 2 \\
		${\rm H12}$ & ${\rm 2 H} \rightarrow  {\rm H^{+}} + {\rm e} +  {\rm H}$ & $k_{\rm H12} = 1.7 \times 10^{-4} k_{\rm H11}$ & 2  \\
		${\rm H13}$ & ${\rm H^{-}} + {\rm e} \rightarrow  {\rm H} + 2 {\rm e}$ &  $k_{\rm H13} = \exp [ - 18.01849334 + 2.3608522 \times  \left( \ln T \left({\rm ev} \right) \right)-0.28274430 \left( \ln T \left({\rm ev} \right) \right)^{2} +1.62331664 \times 10^{-2}  \left( \ln T \left({\rm ev} \right) \right)^{3}  $& 7 \\
		&& \hspace{0.5cm} $-3.36501203 \times 10^{-2}   \left( \ln T \left({\rm ev} \right) \right)^{4} +1.17832978 \times 10^{-2}  \left( \ln T \left({\rm ev} \right) \right)^{5}  -1.65619470 \times 10^{-3}  \left( \ln T \left({\rm ev} \right) \right)^{6} $ & \\
		& & \hspace{0.5cm} $+1.06827520 \times 10^{-4}  \left( \ln T \left({\rm ev} \right) \right)^{7} - 2.63128581 \times 10^{-6}  \left( \ln T \left({\rm ev} \right) \right)^{8} $ & \\
		${\rm H14}$ & ${\rm H^{-}} + {\rm H^{+}} \rightarrow  2 {\rm H} $ & $k_{\rm H14} = 6.3 \times 10^{-8} + 5.7 \times 10^{-6} / T^{0.5} - 9.2 \times 10^{-11} T^{0.5} + 4.4 \times 10^{-13} T$ & 2\\

        ${\rm RH1}$ & ${\rm H} + {h \nu} \rightarrow {\rm H^{+} + {\rm e}}$ & $R_{1}$ & Eq. \eqref{0816.HI} \\
        ${\rm RH2}$ & ${\rm H_{2}} + {h \nu} \rightarrow {\rm 2 H}$ & $R_{2}$& Eq. \eqref{0816.H2} \\
 		\hline
 	\end{tabular}
	 \begin{minipage}{1\hsize}
 	 References. (1) \citet{2006agna.book.....O}; (2)\citet{1998A&A...335..403G}; (3)\citet{1987ApJ...318...32S}; (4)\citet{2013ApJ...773L..25F}; (5)\citet{1983ApJ...271..632P}; (6) \citet{2010Sci...329...69K}; (7)\citet{1997NewA....2..181A}; (8)\citet{1985ApJ...291..722T}
	 \end{minipage}
 \end{table*}

 \subsection{Thermal Processes}\label{thermal_process}
 In Table \ref{thermal_process_t}, we summarize the heating and cooling processes incorporated in our simulations. In what follows, we describe the details of each process.

 \begin{table*}
 \caption{Thermal Processes}
 \label{thermal_process_t}
 \begin{tabular}{cllc}
 	\hline \hline
  Number & Process & Rate (${\rm erg \, cm^{-3} \, s^{-1}}$) & Reference \\
  \hline \hline
  \multicolumn{4}{|c|}{Heating}\\ \hline
  1 & ${\rm H_2}$ formation & $\Gamma_{1} = [ \, 3.53 (1+ n_{\rm cr} / n_{\rm H})^{-1} k_{\rm H3} n({\rm H}) n({\rm H^{-}}) + 4.48 (1+ n_{\rm cr} / n_{\rm H})^{-1} k_{\rm H5} n^{3}({\rm H})  $ & \\
  & & \hspace{7mm} $+ \, \left( 0.2 + 4.2 (1+ n_{\rm cr} / n_{\rm H})^{-1} \right) k_{\rm H8} n^{2}({\rm H})] ~ \rm{eV}  $ &  \\
  & &  $n_{\rm cr} = 10^{6}T^{-1/2}/ \left\{ 1.6 n({\rm H}) / n_{\rm H} \exp [-(400/T)^{2}] + 1.4 n({\rm H_2})/n_{\rm H} \exp[-1200/(T+1200)]  \right\} ~ {\rm cm^{-3}} $ & 1,2\\
  2 & H photoionization & $\Gamma_{2}$ & Eq. \eqref{0816.6} \\

 \hline \multicolumn{4}{|c|}{Cooling}\\  \hline
 1 & ${\rm H_{2}}$ dissociation & $\Lambda_{1} = 4.48 \, [ \, k_{\rm H4} n({\rm H_2}) n({\rm H}) + k_{\rm H7} n^{2}({\rm H_2}) + k_{\rm H11} n({\rm H}) n({\rm e}) \, ] \, {\rm eV}$ & 1,2 \\
 2 & H ionization & $\Lambda_{2} = 13.6 \, [ \, k_{\rm H9} n({\rm H}) n({\rm e}) + k_{\rm H12} n^{2}({\rm H}) \,] \, {\rm eV} $ & 1,2 \\
 3 & ${\rm H^{-}}$ free-bound emission & $\Lambda_{3} = 0.755 \, k_{\rm H2} n({\rm H}) n({\rm e}) \, {\rm eV}$ & 6 \\
 4 & H excitation & $\Lambda_{4} = 7.5 \times 10^{-19} \left[ 1 + (T / 10^{5} ~ {\rm K})^{1/2} \right]^{-1} \exp[-118348/T]n({\rm H})n({\rm e})$ & 3 \\
 5 & H recombination & $\Lambda_{5} = 2/3 k_{\rm B} T k_{\rm H1} n({\rm H^+}) n({\rm e})$ & 4 \\
 6 & Free-free & $\Lambda_{6} = 1.426 \times 10^{-27} T^{-1/2} g_{ff}(T;1) n({\rm H}) n({\rm e})$ & 5 \\
 & & $g_{ff} (T;Z) = 0.79464 + 0.1243 \log10(T/Z^{2})$ \hspace{5mm} $T/Z^{2} < 32000 ~ {\rm K}$ & \\
 & & \hspace{13.5mm}$ = 2.13164 - 0.1240 \log10(T/Z^{2})$ \hspace{5mm} $T/Z^{2} > 32000 ~ {\rm K}$ & \\
 7 & Compton & $\Lambda_{7} = 1.017 \times 10^{-37} T_{\rm CMB}^{4} (T - T_{\rm CMB}) n({\rm e})$ & 3 \\
 8 & Gas-grain heat transfer & $\Lambda_{8} = 5.83 \times 10^{-8} n_{\rm H} \rho \left( T / 10^{3} ~ {\rm K} \right)^{1/2} \left[ 1 - 0.8 \exp(- 75 ~ {\rm K} / T) \right] \left( T - T_{\rm gr} \right) \left( Z / Z_{\odot} \right) \epsilon_{\rm d} $ & 1,2 \\
 9 & Line cooling & $\Lambda_{9}  =\Lambda_{\rm H_{2}} + \Lambda_{\rm CII} + \Lambda_{\rm OI} + \Lambda_{\rm OII} + \Lambda_{\rm OIII} $ & \\

 \hline \hline
 \end{tabular}
  \begin{minipage}{1\hsize}
  References. (1) \citet{1979ApJS...41..555H} (2) \citet{2000ApJ...534..809O} (3) \citet{1992ApJS...78..341C} (4)\citet{1978ppim.book.....S} (5) \citet{1987ApJ...318...32S} (6) \citet{2016ApJ...824..119H}
  \end{minipage}
\end{table*}

\subsubsection{line cooling}
As discussed in Section \ref{chem_therm}, we consider cooling of ${\rm H_{2}}$ rovibrational transitions, atomic fine structure transitions of C{\sc ii} and O{\sc i}, collisional excitation of O{\sc ii} and O{\sc iii}:
 \begin{eqnarray}
	 \Lambda_{9}  =\Lambda_{\rm H_{2}} + \Lambda_{\rm CII} + \Lambda_{\rm OI} + \Lambda_{\rm OII} + \Lambda_{\rm OIII}. \label{0813.41}
\end{eqnarray}
For cooling of $\rm H_2$ rovibrational transitions, we use the cooling function obtained by \citet{1998A&A...335..403G} and the escape probability given by \citet{2018MNRAS.473.4754F}.
We calculate the cooling rate of atomic transitions as \citep{2000ApJ...534..809O}:
\begin{eqnarray}
\Lambda_{\rm X} = \frac{1}{\rho} \sum_{(i \rightarrow j)} n_{i} A_{ij} \epsilon_{ij} h \nu_{ij}, \label{0815.1}
\end{eqnarray}
where the label $X$ represents C{\sc ii}, O{\sc i}, O{\sc ii} and O{\sc iii}, and the labels $i$ and $j$ correspond with the energy levels.
In Equation \eqref{0815.1}, $n_{i}$ and $A_{ij}$ are the number density at the energy level $i$ and the Einstein coefficient of energy transition $i \rightarrow j$.
The number density of each energy level is obtained by the solving statistical equilibrium equation as 
\begin{eqnarray}
n_{i}  \sum_{j \neq i} q_{ij} = \sum_{j \neq i} n_{j} q_{ji}, \label{0815.2}
\end{eqnarray}
where $q_{ij}$ represents the transition rate between the energy level $i \rightarrow j$: 
\begin{eqnarray}
  q_{ij} = \begin{cases}
    A_{ij} \epsilon_{ij} + \sum_{k} \gamma_{ij}^{k} n_{k} & {\rm for} \hspace{3mm} i > j \\
     \sum_{k} \gamma_{ij}^{k} n_{k}  & {\rm for} \hspace{3mm}  i < j,
  \end{cases} \label{0815.3}
\end{eqnarray}
In Equation \eqref{0815.3}, $\gamma_{ij}^{k}$ is the collisional excitation/de-excitation rates.
Same as \citet{2018ApJ...857...57N}, we use the rates of O{\sc i} and C{\sc ii} given by \citet{1989ApJ...342..306H} and \citet{2006ApJ...643...26S}.
For O{\sc ii} and O{\sc iii}, we use the rates of \citet{2006agna.book.....O} and \citet{2011piim.book.....D}.  
In Equation \eqref{0813.41}, we use the formula of  escape probability $\epsilon_{ij}$ of \citet{1983ApJ...275..145T} in the same way as \citet{2000ApJ...534..809O}.
 
 \subsubsection{OII \& OIII abundance in H{\sc ii} regions}
\label{ssec:oxy}

 We estimate O{\sc ii} and O{\sc iii} abundance assuming the balance between ionization of O{\sc ii} and recombination of O{\sc iii} as
 \begin{eqnarray}
	n_{\rm OII} \left( R_{\rm OII} + n_{\rm e} \xi_{\rm coll} \right) = n_{\rm OIII} n_{\rm e} \alpha_{\rm OIII}. \label{A2.2.1}
\end{eqnarray}
The recombination rate $\alpha_{\rm OIII}$ is given by \citet{1983A&A...126...75N} and \citet{1991A&A...251..680P}, and the collisional ionization rate $\xi_{\rm col}$ is obtained by \citet{1997ADNDT..65....1V}.
The photoionization rate of O{\sc II} is estimated as 
\begin{eqnarray}
  R_{\rm OII} = \int^{\infty}_{\nu_{\rm OII}} d \nu \sigma_{\rm OII}(\nu) \frac{F_{\nu}}{h \nu}, \label{0816.2}
\end{eqnarray}
where $\sigma_{\rm O{\sc II}}$ and $\nu_{\rm O{\sc II}}$ are the cross section and ionization limit frequency of $\nu_{\rm O{\sc II}}$ \citep{2006agna.book.....O}.

\subsubsection{Hydrogen photoionization and photodissociation}
\label{photoheat}

The photoionization rate $R_1$ and the photoheating rate $\Gamma_2$ is estimated as
\begin{eqnarray}
  R_{1} = y_{\rm HI} \int^{\infty}_{\nu_{l}} d \nu \sigma_{\rm HI}(\nu) \frac{F_{\nu}}{h \nu}, \label{0816.HI}
\end{eqnarray}
\begin{eqnarray}
  \Gamma_{2} = \frac{1}{\rho} n_{\rm HI} \int^{\infty}_{\nu_{l}} d \nu \sigma_{\rm HI}(\nu) F_{\nu} \frac{h\left(\nu - \nu_{\rm L} \right)}{h \nu}. \label{0816.6}
\end{eqnarray}
The photodissociation rate of $\rm H_2$ is estimated as \citep{1996ApJ...468..269D, 2018ApJ...857...57N}
\begin{eqnarray}
  R_{2} = f_{\rm shield}\left(N_{\rm H_2} \right) e^{-\tau_{\rm d, fuv}} I_{\rm diss} n_{\rm H_2}, \label{0816.H2}
\end{eqnarray}
where $e^{-\tau_{\rm d, fuv}}$ is the optical depth of dust grains for FUV photons and $I_{\rm diss} \simeq 4 \times 10^{-11} G_{\rm FUV} ~ {\rm s^{-1}}$ represents the photodissociation rate without self-shielding.
Here, $G_{\rm FUV}$ is estimated as $G_{\rm FUV} = L_{\rm FUV}/(4 \pi r^2 F_{\rm ISRF})$ where $L_{\rm FUV}$ is luminosity of FUV and $F_{\rm ISRF} = 1.6 \times 10^{-3} ~ {\rm erg \, cm^{-2} \, s^{-1}}$. 
We consider the self-shielding of $\rm H_2$ and the self-shielding rate $f_{\rm shield}$ is estimated as 
\begin{eqnarray}
	f_{\rm shield} = \begin{cases}
	1 \hspace{17.5mm} \left( N_{\rm H_{2}} < N_{1} \right) \\
	\left(N_{\rm H_{2}}/ N_{1} \right)^{-3/4} \left(  N_{1} < N_{\rm H_{2}} < N_{2} \right) \\
	0 \hspace{17.5mm} \left( N_{\rm H_{2}} > N_{2} \right), \label{0816.8}
	\end{cases}
\end{eqnarray}
where $N_{\rm H_{2}}$ is the column density of $\rm H_2$ molecules, $N_{1} = 10^{14}~{\rm cm^{-2}}$, and $N_{2} = 10^{22}~{\rm cm^{-2}}$ \citep[e.g.,][]{2011Sci...334.1250H}.

\subsubsection{Dust grains}
\label{tdust}

We estimate the dust temperature assuming the energy balance among (1) dust thermal emission, (2) absorption of the stellar direct light, (3) absorption of the diffuse light, and (4) energy transport from the gas.
We use the opacity table of \citet{1993ApJ...402..441L} and the energy transport rate between dust and gas given as
\begin{align}
    \Lambda_{8} = 5.83 \times 10^{-8} n_{\rm H} \left( \frac{T}{10^3~\rm{K}}\right)^{1/2} \left[ 1 - 0.8 \exp \left( \frac{ -75 {\rm K}}{  T}  \right) \right] \left( T - T_{\rm gr} \right)
\label{0816.9}
\end{align}
 \citep[e.g.,][]{1979ApJS...41..555H,2005ApJ...626..627O}.
We also consider the sublimation of dust grains above a threshold temperature $T_{\rm vap} = g \rho ^{\beta}$, where $g = 2000$  and $\beta = 0.0195$ \citep{2005A&A...438..899I}. When the dust temperature exceeds this value, we decrease the dust abundance over a local dynamical timescale.

\section{Convergence tests}
\label{resol}

\begin{figure}
 \begin{center}
 \includegraphics[width=\columnwidth]{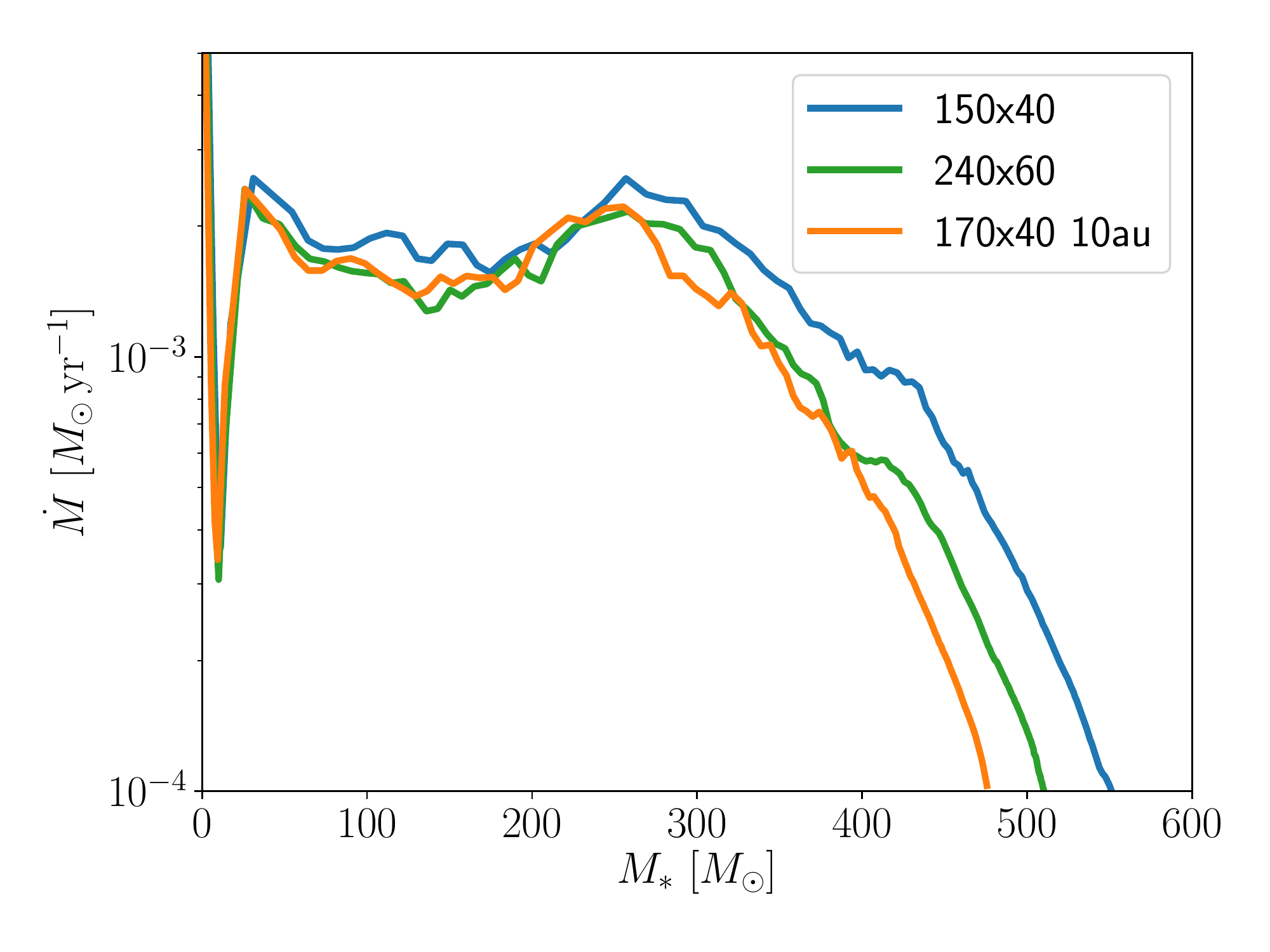}
 \end{center}
 \caption{Evolution of the mass accretion rates at $Z=0$ for the fiducial case Prm (blue), the case with the higher resolution (Prm\_highres, green), and the case with the smaller sink of $10~{\rm au}$ (Prm\_rsink10au, orange).}
 \label{resol41}
 \end{figure}

In order to examine numerical convergence, we here perform additional 2D simulations for which we vary the spatial resolution and size of the sink cell. 
We use the snapshot at the end of the early collapse stage at $Z=0$ as the initial condition for those simulations, as in case Prm described in the main part. 
Figure \ref{resol41} shows the accretion histories for the cases with the default resolution ($N_{\rm r} \cdot N_{\theta}) = (150 \cdot 40$), the higher resolution ($N_{\rm r} \cdot N_{\theta}) = (240 \cdot 60$), and the smaller sink radius $10~{\rm au}$ with ($N_{\rm r} \cdot N_{\theta}) = (170 \cdot 40$), for which the cell sizes are the same as for the default resolution (models Prm\_highres and Prm\_rsink10au in Table 1).
In comparison to the fiducial model Prm, the final stellar masses decrease in the other cases, but the differences are less than a few $\times$ 10 \%.
The decreasing rate of the mass accretion rate is almost the same after the protostar accretes gas of $400~M_{\odot}$.
Owing to high computational costs, we decide that the resolution of $(N_{\rm r} \cdot N_{\theta}) = (150 \cdot 40)$ and the sink radius of $30~{\rm au}$ are sufficient to investigate massive star formation in our framework.

\section{Comparison of 3D and 2D simulation data}
\label{map_3Dto2D}
 \begin{figure}
 \begin{center}
 \includegraphics[width=\columnwidth]{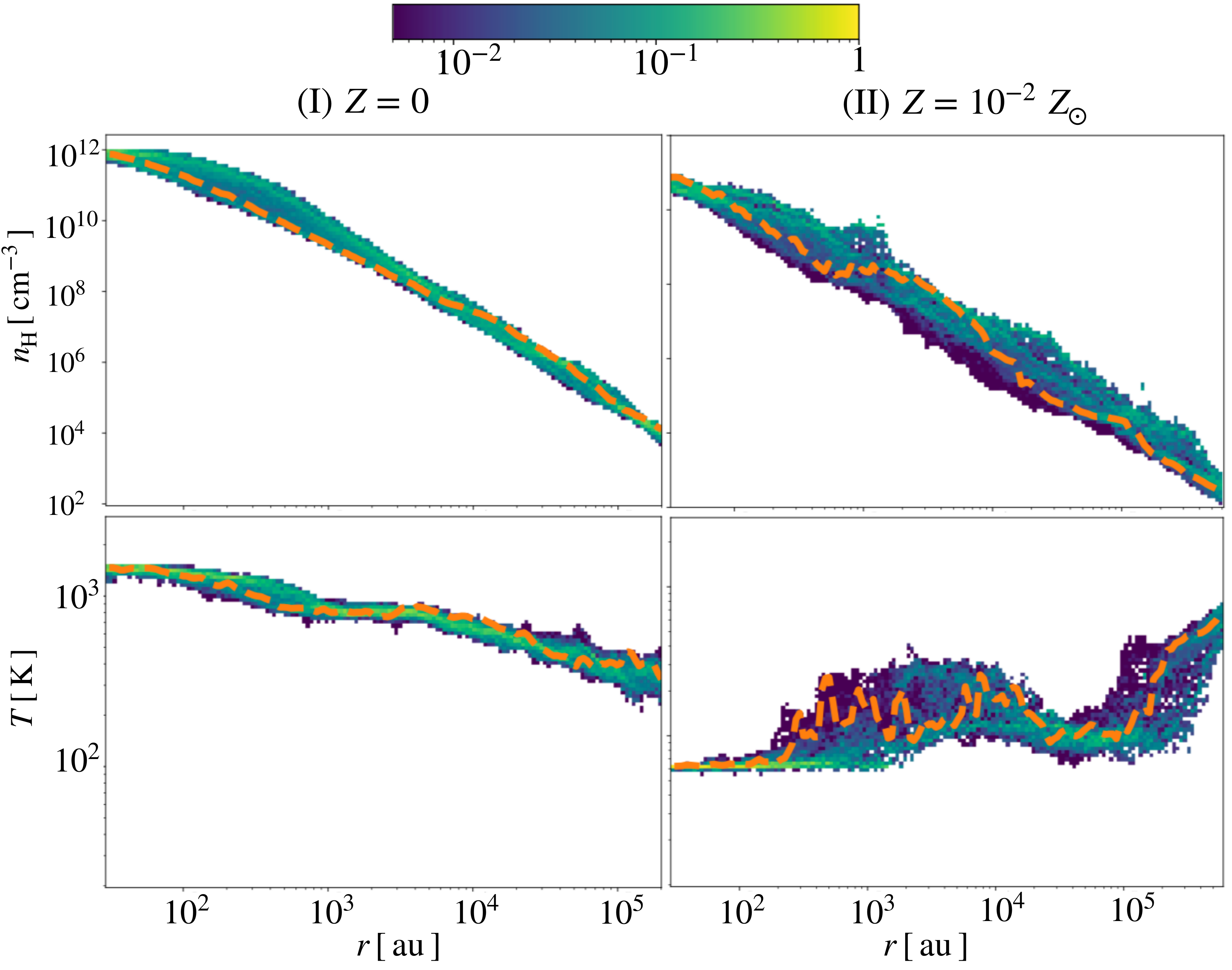}
 \end{center}
 \caption{Differences between the original 3D data and the mapped 2D data. Top and bottom panels show the temperature and density distributions on the disk midplane in the cases with $Z=0$ (left column) and $Z=10^{-2}Z_{\odot}$ (right column). The colors represent the mass fraction of 3D cells that have the corresponding temperature and density. The orange dashed lines represent the values used in the 2D simulation.}
\label{dis_rhoT_ZP}
 \end{figure}
 \begin{figure}
 \begin{center}
 \includegraphics[width=\columnwidth]{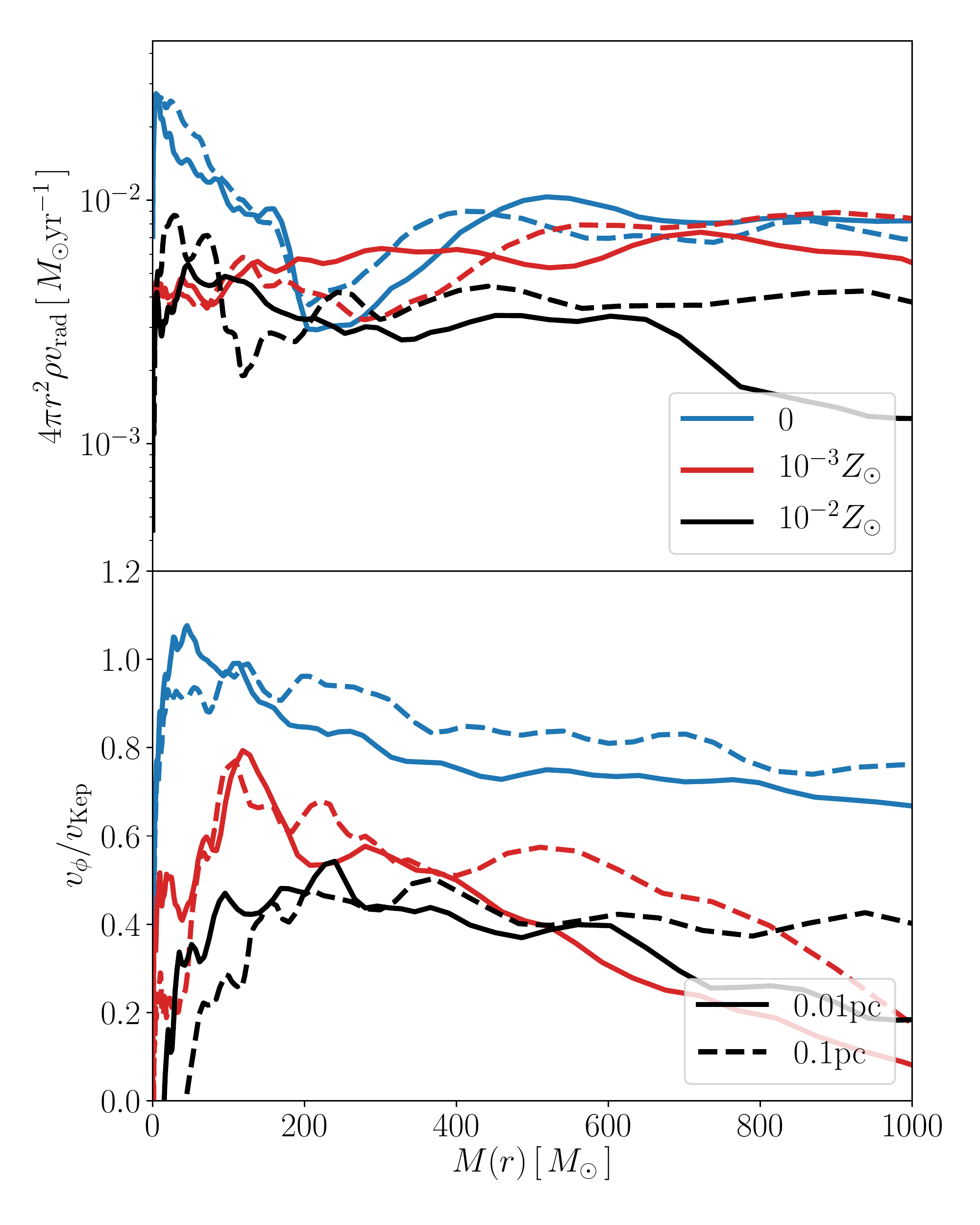}
 \end{center}
 \caption{
 Same as Figure \ref{zu_init1} but for cases with different sizes of the central region 
 set for determining the rotational axis. The top and bottom panels show the expected mass inflow rate and rotational velocity distributions at the end of the collapse phase. 
 The solid and dashed lines represent the cases where we calculate the angular momentum vector in the $0.01$ and $0.1~{\rm pc}$ central regions.
 }
 \label{hikaku_zu1}
 \end{figure}
When we generate the initial conditions for our simulations of protostellar accretion, we convert the original 3D data to 2D data in the polar coordinates, as described in Section \ref{initial_conditions}. 
To do this, we simply use physical values in a plane.
We study how our strategy 
affects the simulations by estimating the information lost through the data mapping process. Figure \ref{dis_rhoT_ZP} show the 
differences of the density and temperature between the 3D and 2D data. 
In the case with $Z=0$, the 3D data shows small scatter around the averaged 2D profiles
at a given radius. The deviations in the azimuthal direction are small because the cloud has  nearly spherical morphology, as shown in Figure \ref{snap_shot3D}. 
The gas distribution is well approximated by the axially symmetric profiles. 
Contrastingly, the cloud with $Z=10^{-2}Z_{\odot}$ has filamentary structure that has developed during the collapse phase owing to efficient metal cooling. 
Such non-axisymmetric structure causes deviations from mapped values, as indicated by the spread of the 3D data presented in Figure~\ref{dis_rhoT_ZP}.
Specifically, the temperature distribution shows the large scatter at $r \sim 10^3$~AU and our 2D profile does not necessarily trace the typical values. However, the scatter is relatively small for the range of $10^4 - 10^5~{\rm AU}$, 
within which $\sim 100 - 1000~M_\odot$ of the gas is enclosed   (see Fig.~\ref{zu_init1}). Since we mainly consider the stellar radiative feedback that becomes effective after the star accretes $\sim 100~\msun$ of the gas, we expect that the scatters seen in the 3D data do not significantly change our conclusions.

In the mapping procedure, we calculate the angular momentum of the gas contained in the
central $0.01~{\rm pc}$ ($2\times 10^3~{\rm au}$) region in order to fix the rotational axis.
Since our choice of $0.01~{\rm pc}$ is somewhat arbitrary, we study the effect of varying the radial extent. We compare the expected mass inflow rate and the rotational velocity with different choices of $0.01$ and $0.1~{\rm pc}$ in Figure \ref{hikaku_zu1}. In the primordial case with $Z=0$, the cloud is nearly spherical, and the direction of the angular momentum vector does not differ significantly regardless of the size of the region considered. Note that the differences in the radial distribution are also small (Figure \ref{hikaku_zu1}). 
The deviations are more prominent for the cases with higher metallicities $Z=10^{-3}$ and $10^{-2}~Z_\odot$. If the cloud has filamentary structure, the angular momentum vector differs slightly between the cases with $0.01$ and $0.1~{\rm pc}$. Nonetheless, the differences are still only a few $\times 10$~\% in the outer parts of the accretion envelope $M(r) \gtrsim 100~M_\odot$. Therefore, we expect that our results do not crucially depend on the choice of the central region for fixing the rotational axis.

\bsp	
\label{lastpage}
\end{document}